\documentclass[aps,twocolumn,showpacs,floatfix]{revtex4}

\usepackage[usenames]{color} 

\usepackage{amssymb,epsfig}

\usepackage{psfrag}

\renewcommand{\theequation}{\arabic{section}.\arabic{equation}}

\newcommand{\beq}[1]{
\begin{equation}\label{#1}}
\newcommand{\eeq}{\end{equation}}
\newcommand{\bea}[1]{
\begin{eqnarray}\label{#1}}
\newcommand{\eea}{\end{eqnarray}}

\newcommand{\bra}[1]{\left\langle #1 \right|}
\newcommand{\ket}[1]{\left| #1 \right\rangle}

\newcommand{\Gl}[1]{Eq.~(\ref{#1})}

\newcommand{\al}{\alpha}
\newcommand{\be}{\beta}  
\newcommand{\ep}{\varepsilon}
\newcommand{\ga}{\gamma}
\newcommand{\de}{\delta}

\newcommand{\la}{\lambda}

\newcommand{\si}{\sigma}

\newcommand{\dd}{{\rm d}}

\newcommand{\V}{{\cal V}}  
\newcommand{\A}{{\cal A}}

\newcommand{\nn}{\nonumber}

\newcommand{\lash}[1]{\not\! #1 \,}

\psfrag{Q2}{{\tiny $Q^2$}}

\psfrag{pin}{{\scriptsize $\pi^+n$}}

\psfrag{pip}{{\scriptsize $\pi^0p$}}

\psfrag{pin}{{\scriptsize $\pi^+n$}}

\psfrag{pip}{{\scriptsize $\pi^0p$}}

\psfrag{G1GA}{{\scriptsize $\sqrt{2}Q^2G_1^{\pi N}/(G_A m_N^2)$}}

\psfrag{G2GA}{{\scriptsize $\sqrt{2}|G_2^{\pi N}|/G_A$}}

\begin{document}



\title{Deep Inelastic Pion Electroproduction at Threshold}

\author{V. M. Braun$^1$, D. Yu. Ivanov$^2$, A. Lenz$^1$ and A. Peters$^1$\\}

\affiliation{
{}$^1$ Institut f\"ur Theoretische Physik, Universit\"at
          Regensburg, D-93040 Regensburg, Germany \\
{}$^2$ Sobolev Institute of Mathematics, 630090 Novosibirsk, Russia}

\date{\today}

\begin{abstract}
We consider the cross section of the deep inelastic pion electroproduction on a proton target 
at threshold for $Q^2$ in the region $5-10$~GeV$^2$. The corresponding amplitudes are described in terms  
of two form factors which we calculate using light cone sum rules (LCSR) to 
leading order in QCD and including higher twist corrections. Our results suggest a considerable change 
from a small $Q^2$ region that can be treated in the soft pion limit using current algebra.
In particular, we obtain a $\pi^0$ to $\pi^+$ producton ratio of order 1/3 and significant nucleon helicity-flip 
contributions. 
\end{abstract}






\maketitle


\section{Introduction}
\setcounter{equation}{0}
In recent years there has been increasing attention to hard exclusive processes involving 
emission of soft pions in the final state. One reason is that such processes often provide 
the main background to the simpler reactions that one wants to study, and one would like 
to estimate the background as precise as possible. The main motivation is, however, different.
It turns out that hard production of soft pions is interesting in its own right and can provide 
one with new insights in the hadron (nucleon) structure. The novel physical feature is in this
case the presence of three disparate scales: the large momentum transfer $Q$, the QCD scale 
$\Lambda = \Lambda_{\rm QCD}$, and the pion mass $m_\pi$ which goes to zero in the chiral limit, so 
that $Q \gg \Lambda_{\rm QCD} \gg m_\pi$. Moreover, the limits $m_\pi\to 0$ and $Q\to\infty$
do not commute. If, parametrically,  $Q^2 \ll \Lambda^3/m_\pi$ then the standard 
approach based on chiral symmetry and current algebra is applicable.
In this way the celebrated soft pion theorems arise \cite{Kroll:1953vq}:  the amplitudes
with pion emission are calculated in terms of the amplitudes without pions induced by a chirally-rotated 
current, and pion emission from the external hadron lines. There exists vast literature on this topic, see e.g. 
\cite{Nambu:1997wb,Vainshtein:1972ih,Scherer:1991cy}, extending the original Kroll-Ruderman prediction to include
corrections in the pion mass and/or the pion (photon) momentum.

In the opposite limit  $Q^2 \gg \Lambda^3/m_\pi$ these  techniques do not apply, and the accepted
general philosophy is to use QCD factorization to separate contributions of large momenta of the order of the 
hard scale $Q$ in perturbatively calculable ``coefficient functions'' in front of nonperturbative
matrix elements that contain contributions of smaller scales. The methods of current algebra
(or chiral perturbation theory) can then be used to simplify the matrix elements rather than
the full physical amplitude in question. The present study is fuelled by the observation made in Ref.~\cite{PPS01}
that in the asymptotic $Q\to\infty$ limit the pion electroproduction can be described in the framework of the usual 
pQCD factorization formalism for exclusive processes \cite{Chernyak:1977as,Radyushkin:1977gp,Lepage:1979zb}
and involve an overlap integral of chirally rotated nucleon light-cone distribution amplitudes (DAs). 
The particular combination of the DAs involved in the pion electroproduction turns out to be different from that 
in the electromagnetic (or weak) elastic nucleon form factors, so that different components of DAs 
potentially could be separated from the comparison. This is interesting, since nucleon light-cone DAs remain 
to be poorly known. Similar ideas have been applied in a somewhat different context in 
Refs.~\cite{Pire:2005ax,Lansberg:2006nv}.

The essential requirement for the applicability  of the pQCD factorization is a high virtuality 
of the exchanged gluons and also of the quarks inside the short distance subprocess. 
The main problem is a numerical suppression of each hard gluon exchange by the $\alpha_s/\pi$ factor which is a 
standard perturbation theory penalty for each extra loop. 
If, say,  $\alpha_s/\pi \sim 0.1$, the pQCD 
contribution to baryon form factors is suppressed by a factor of 
100 compared to the purely soft term.  As the result, 
the onset of the perturbative regime is postponed to very large momentum transfers since 
the factorizable pQCD contribution  has to win over (nonfactorizable) nonperturbative effects 
that are suppressed by  extra powers of $1/Q^2$ but do not involve small coefficients.  
There is a growing consensus that such ``soft'' contributions play the dominant role at present 
energies. Indeed, it is known for a long time that the use of 
QCD-motivated models for the  wave functions allows one to
obtain, without much effort,  soft contributions comparable in size 
to  experimentally observed values (see, e.g.~\cite{Isgur:1984jm,Isgur:1988iw,Kroll:1995pv}). 
A current fashion \cite{Radyushkin:1998rt,Diehl:1998kh} is to  use the concept of  
generalized parton distributions  (GPDs) to describe/parametrize soft contributions in various 
exclusive reactions, see  \cite{Goeke:2001tz,Diehl:2003ny,Belitsky:2005qn} for recent reviews,
and the models of GPDs usually are chosen such that the experimental data on form factors 
are described by the soft contributions alone, cf. Refs.~\cite{Belitsky:2003nz,Diehl:2004cx,Guidal:2004nd}. 
The soft pion electroproduction was considered in this framework in \cite{Guichon:2003ah,Polyakov:2006dd}.
A subtle  point for these semi-phenomenological 
approaches is to avoid double counting of hard rescattering 
contributions  ``hidden'' in the model-dependent hadron wave functions
or GPD parametrizations.  
   
Another approach to the calculation of form factors for moderately large $Q^2$ is based on the 
light-cone  sum rules (LCSR) \cite{Balitsky:1989ry,Chernyak:1990ag}.  
This technique is attractive because in LCSRs  ``soft'' contributions to the form factors are calculated in 
terms of the same DAs that enter the pQCD calculation and there is no double counting. Thus, the LCSRs provide one with the most 
direct relation of the hadron form factors and distribution amplitudes that is available at present, 
with no other nonperturbative parameters.  

\begin{figure}[ht]
\centerline{\epsfxsize5cm\epsffile{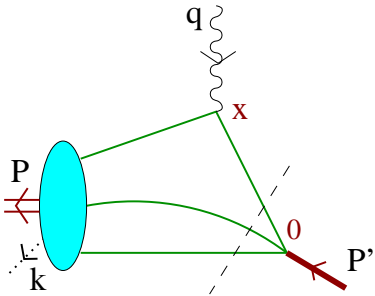}}
\caption{\small
Schematic structure of the light-cone sum rule for pion electroproduction
}
\label{figsum}
\end{figure}

In this paper, we suggest to use light-cone sum rules to calculate the amplitudes of pion electroproduction. 
The basic object of the LCSR approach is the 
correlation function $$\int\! dx\, e^{-iqx}\langle N(P)\pi(k)| T \{ j(x) \eta (0)\} | 0 \rangle $$ in which
$j$  represents the electromagnetic  probe and $\eta$
is a suitable operator with nucleon quantum numbers.    
 The final state nucleon and the pion are explicitly represented by the state vector 
 $\langle  N(P) \pi(k)| $, see a schematic representation in Fig.~\ref{figsum}.
 When both  the momentum transfer  $Q^2$ and 
 the momentum $(P')^2 = (P-q+k)^2$ flowing in the $\eta$ vertex are large and negative,
 the asymptotics of the correlation function is governed by the light-cone kinematics $x^2\to 0$ and
 can be studied using the operator product expansion (OPE)   
$T \{ j(x) \eta(0) \} \sim \sum C_i(x) {\cal O}_i(0)$ on the 
light-cone $x^2=0$.   The  $x^2$-singularity  of a particular perturbatively calculable
short-distance factor  $C_i(x)$  is determined by the twist of the relevant
composite operator ${\cal O}_i$, whose matrix element $\langle N(P)\pi(k)|  {\cal O}_i(0)|0 \rangle $
is given by an appropriate moment of the (complex conjugated) pion-nucleon generalized DA. 
Using current algebra and chiral symmetry these matrix element can be reduced to the usual nucleon DAs. 
Next, one can represent the answer in form of the dispersion integral in $(P')^2$ and define the nucleon contribution
by the cutoff in the invariant mass of the three-quark system, 
the so-called interval of duality $s_0$ (or continuum threshold).
The main role of the interval of duality is that it does not allow large momenta $|k^2| > s_0$ to flow through the 
 $\eta$-vertex; to the lowest order $O(\alpha_s^0)$ one obtains a purely soft    
contribution to the form factor as a sum of terms ordered by twist of the relevant operators and
hence including both the leading- and the higher-twist nucleon DAs. Note that, in difference to the hard mechanism, the 
contribution of higher-twist DAs is only suppressed by powers of 
$|(P')^2|\sim 1-2$~GeV$^2$ (which is translated to the suppression
by powers of the Borel parameter after applying the usual QCD sum rule machinery), but not by powers of $Q^2$. This feature is
in agreement with the common wisdom that soft contributions are not constrained to small transverse separations.  

The LCSR expansion also  contains terms  
generating the asymptotic pQCD contributions. They   appear 
at proper order in $\alpha_s$, i.e., in  the $O(\alpha_s)$ term for the
pion form factor, at the $\alpha_s^2$ order for the nucleon form factors, etc. 
In the pion case, it was explicitly demonstrated 
\cite{Braun:1999uj,Bijnens:2002mg} that the contribution of hard  
rescattering is correctly reproduced in the LCSR   
approach as a part of the $O(\alpha_s)$ correction.
It should be noted that  the  diagrams of LCSR that 
contain the ``hard'' pQCD  contributions also possess ``soft'' parts,
i.e., one should perform  a separation  of ``hard'' and ``soft''
terms inside each diagram.  As a result, 
the distinction between ``hard'' and ``soft'' contributions appears to 
be scale- and scheme-dependent \cite{Braun:1999uj}. 
During the  last years there have been numerous applications of LCSRs  
to mesons, see \cite{Braun:1997kw,Colangelo:2000dp} for a review.
{}Following the work \cite{Braun:2001tj} nucleon electromagnetic form factors 
were further considered in this framework in Refs.~\cite{Lenz:2003tq,Wang:2006uv,Wang:2006su,Braun:2006hz} 
and the weak decays $\Lambda_b\to p\ell\nu_\ell$, $\Lambda_c\to \Lambda \ell\nu_\ell$ etc. 
in \cite{Huang:2004vf,Huang:2006ny,Wang:2006yz}. A generalization to the $N\gamma\Delta$ transition form factor
was worked out in \cite{Braun:2005be}.

The presentation is organized as follows. Section~2 is mainly introductory. We introduce here the 
necessary notations, explain the kinematics and define the two independent invariant 
amplitudes (generalized form factors) that are relevant for the pion electroproduction at threshold.
The main nonperturbative input in our calculations is provided by the generalized pion-nucleon distribution
amplitudes which are calculated using current algebra in Section~3. 
The light-cone sum rules are constructed in Section~4. The numerical analysis is 
done in Section~5 which contains our main results.
Finally, Section~6 contains a short summary and conclusions.

\section{Kinematics and Notations}
\setcounter{equation}{0}
In this paper we consider pion electroproduction from a proton target
\bea{piprod}
  e(l')+p(P') &\to& e(l) + \pi^+(k) + n(P)\,,
\nonumber\\
  e(l')+p(P') &\to& e(l) + \pi^0(k) + p(P)\,.
\eea
Basic kinematic variables are 
\begin{eqnarray}
&&q=l'-l\, , \quad s=(l'+P')^2, \quad W^2=(k+P)^2 \,, 
\nonumber\\
&& q^2=-Q^2 \, , \quad P'^2=P^2=m_N^2 \, , \quad k^2=m^2_\pi\,, 
\nonumber \\ 
&& y=\frac{
P'\cdot q}{ P'\cdot l'}=\frac{
W^2+Q^2-m_N^2}{ s-m_N^2}\,.
\end{eqnarray}
The identification of the momenta is clear from Eq.~(\ref{piprod});
$m_N$ is the nucleon and $m_\pi$ the pion mass, respectively.
In what follows we neglect the electron mass and the difference of proton and neutron masses.

The differential cross section is equal to
\beq{difsec} \frac{d\sigma}{dy d Q^2 d\phi} =
\frac{1}{(2\pi)^5}\frac{|\mathcal M|^2}{64(s-m_N^2)}\beta(W)
 d\phi_\pi d(\cos\theta)\,.
\eeq
Here ${\mathcal M}$ is the amplitude:
\beq{amplitude}
{\mathcal M} =\frac{4\pi \alpha_{em}}{q^2}\bar u(l)\gamma_\mu u(l')
\langle N \pi |j_\mu^{em}(0)| p\rangle\,,
\eeq
$\theta$ and $ \phi_\pi$ are the polar and azimuthal angles of the pion in the 
final nucleon-pion c.m. frame, respectively, $\phi$ is the azimuth of the scattered electron 
in the same frame, the electromagnetic current is defined as
\beq{jmu}
 j^{\mathrm{em}}_\mu (x) = e_u\bar u(x) \gamma_\mu u(x) + e_d\bar d(x) \gamma_\mu d(x)\,,
\eeq
and $\beta(W)$ is the kinematic factor related to the c.m.s. momentum of the 
subprocess $\gamma^*(q)+p(P') \to \pi(k)+N(P)$ in the final state:
\bea{beta}
&&\vec k^2_f = \frac{W^2}{4}\left( 1-
\frac{(m_N\!+\!m_\pi)^2}{W^2}\right) \left( 1-
\frac{(m_N\!-\!m_\pi)^2}{W^2}\right),
\nonumber\\
&&\beta(W)=\frac{2|\vec k_f|}{W}\,. 
\eea
Alternatively, one could  use the Mandelstam $t$-variable (of the $\gamma^*p \to \pi N$ subprocess) 
$t=(P'-P)^2$:\\[-3mm] 
\beq{dt}
dt=2|\vec k_i||\vec k_f|d(\cos \theta) \,,
\eeq
where $\vec k_i$ is the c.m.s. momentum in the initial state:
\beq{cmsi} 
\vec k^2_i = \frac{W^2}{4}\left( 1-
2\frac{m_N^2-Q^2}{W^2} + \frac{(m_N^2+Q^2)^2}{W^4}\right). 
\eeq 

We will be interested in the pion production exactly at threshold, in which case
the pion momentum is simply proportional to that of the final state nucleon:
\bea{k}
   && k_\mu = \delta\, P_{\mu}\,,\qquad \delta = m_\pi/m_N \simeq 0.15\,,
\eea
and assume that the photon virtuality is large, $Q^2 \gg \Lambda_{\mathrm QCD}^2$.
The conditions of Lorentz covariance and electromagnetic current 
conservation imply that the amplitudes at threshold can be parametrized 
in terms of two generalized form factors which we define as
\begin{widetext}
\beq{def:gff}
 \langle N(P)\pi(k) |j_\mu^{em}(0)| p(P')\rangle = - \frac{i}{f_\pi} \bar N(P)\gamma_5
  \left\{\left(\gamma_\mu q^2 - q_\mu \!\not\! q\right) \frac{1}{m_N^2} G_1^{\pi N}(Q^2) 
    - \frac{i \sigma_{\mu\nu}q^\nu}{2m_N} G_2^{\pi N}(Q^2)\right\}N(P')\,.
\eeq   
\end{widetext}
Hereafter $N(P)$ is the usual Dirac $u$-spinor, $f_\pi = 93$~MeV. 
For example, the $S$-wave contribution to the structure functions of the 
total deep-inelastic (DIS) cross section from unpolarized protons close to the pion-nucleon threshold 
is given in terms of the generalized form factors as:
\begin{widetext}
\bea{DIS} F_1&=&\frac{\beta(W)}{(4\pi
f_\pi)^2}\frac{Q^2+(2m_N+m_\pi)^2}{2 m_N^3(m_N+m_\pi)} \left(G_1^{\pi N}
Q^2- \frac12 G_2^{\pi N} m_N \, m_\pi\right)^2, 
\nonumber\\
F_2&=&\frac{\beta(W)}{(4\pi
f_\pi)^2}\frac{Q^2(Q^2+m_\pi(2m_N+m_\pi))}{ m_N^3(m_N+m_\pi)}
\left((G_1^{\pi N})^2 Q^2+ \frac14 (G_2^{\pi N})^2 m_N^2 \right). 
\eea   
\end{widetext}
The calculation of $G_1^{\pi N}(Q^2)$ and $G_2^{\pi N}(Q^2)$ in the light-cone sum rule 
approach presents our main goal.

Having in mind the practical construction of light-cone sum rules that involve nucleon DAs,
we define a light-like vector $z_\mu$ by the condition
\beq{z}
       q\cdot z =0\,,\qquad z^2 =0
\eeq
and introduce the second light-like vector 
\bea{vectors}
p_\mu &=& P_\mu  - \frac{1}{2} \, z_\mu \frac{m_N^2}{P\cdot z}\,,~~~~~ p^2=0\,, 
\eea
so that $P \to p$ if the nucleon mass can be neglected, $m_N \to 0$.
The photon momentum can be written as 
\begin{eqnarray}
q_{\mu}=q_{\bot \mu}+ z_{\mu}\frac{P\cdot q}{P\cdot z}\, .
\end{eqnarray}
We also need the projector onto the directions orthogonal to $p$ and $z$,
\begin{equation}
       g^\perp_{\mu\nu} = g_{\mu\nu} -\frac{1}{pz}(p_\mu z_\nu+ p_\nu z_\mu),
\end{equation}
and use the notation
\begin{equation}
    a_z\equiv a_\mu z^\mu, \qquad  a_p\equiv a_\mu p^\mu\,,
\end{equation}
for arbitrary Lorentz vectors $a_\mu$.
In turn, $a_\perp$ denotes the generic component of $a_\mu$ orthogonal to
$z$ and $p$, in particular
\beq{qperp}
   q_{\perp\mu} = q_\mu -\frac{p\cdot q}{p\cdot z} z_\mu\,.
\eeq

We use the standard Bjorken--Drell
convention \cite{BD65} for the metric and the Dirac matrices; in particular,
$\gamma_{5} = i \gamma^{0} \gamma^{1} \gamma^{2} \gamma^{3}$,
and the Levi-Civita tensor $\epsilon_{\mu \nu \lambda \sigma}$
is defined as the totally antisymmetric tensor with $\epsilon_{0123} = 1$.

Assume for a moment that the nucleon moves in the positive 
${\bf e_z}$ direction, then $p^+$ and $z^-$ are the only nonvanishing 
components of $p$ and $z$, respectively. 
The infinite momentum frame can be visualized
as the limit $p^+ \sim Q \to \infty$ with fixed $P\cdot z = p \cdot z \sim 1$ 
where $Q$ is the large scale in the process.
Expanding the matrix element in powers of $1/p^+$ introduces
the power counting in $Q$. In this language,  twist counts the
suppression in powers of $p^+$. Similarly, 
the nucleon spinor $N(P,\la)$ has to be decomposed 
in ``large'' and ``small'' components as
\bea{spinor}
N(P,\la) &=& \frac{1}{2 p\cdot z} \left(\!\not\!{p}\! \!\not\!{z} +
\!\not\!{z}\!\!\not\!{p} \right) N(P,\la)
\nonumber\\&=& N^+(P,\la) + N^-(P,\la)\,,
\eea
where we have introduced two projection operators
\beq{project}
\Lambda^+ = \frac{\!\not\!{p}\! \!\not\!{z}}{2 p\cdot z} \quad ,\quad
\Lambda^- = \frac{\!\not\!{z}\! \!\not\!{p}}{2 p\cdot z}
\eeq
that project onto the ``plus'' and ``minus'' components of the spinor.
Note  the useful relations
\beq{bwgl}
\lash{p} N(P) = m_N N^+(P)\,,\quad \lash{z} N(P) = \frac{2 p \cdot z}{m_N} N^-(P)
\eeq
that are a consequence of the Dirac equation \hfill\break 
$\not\hspace*{-0.10cm}P N(P) = m_N N(P)$.
Using the explicit expressions for $N(P)$ it is easy to see 
that $\Lambda^+N = N^+ \sim \sqrt{p^+}$ while $\Lambda^-N = N^- \sim 1/\sqrt{p^+}$.

Note that all expressions are invariant under the reparametrization $z_\mu \to \alpha z_\mu$
where $\alpha$ is a real  number; we will use this freedom to set $z_\mu$ equal to the 
``minus'' component of the distance between the currents in the operator product.


\section{Pion Nucleon Generalized Distribution Amplitudes}
\setcounter{equation}{0}

\subsection{Leading Twist}
 Our purpose will be to get a complete classification of three-quark generalized distribution 
amplitudes (GDAs) of a $n\pi^+$ and $p\pi^0$ pair in the limit that the momentum of 
the pion relative to that of the nucleon is small which will be a 
generalization of \cite{BFMS00}. This can be done using soft-pion 
theorems, in the spirit of the work \cite{PPS01}.
 In order to develop the necessary formalism  we first consider the leading
twist GDA in some detail. 

\subsubsection{Nucleon Distribution Amplitudes}

We begin with quoting the necessary portion of the definitions and  results 
from \cite{BFMS00}. 

The notion of hadron distribution amplitudes (DAs) in general refers to 
hadron-to-vacuum matrix elements of nonlocal operators build of quark and 
gluon fields at light-like separations. In this paper we will deal with 
the three-quark matrix element 
%
\begin{widetext}
%
\bea{dreiquark}
\bra{0} \epsilon^{ijk} u_\al^{i'}(a_1 z)\left[a_1 z, a_0 z \right]_{i',i}
u_\be^{j'}(a_2 z) \left[a_2 z, a_0 z \right]_{j',j}
d_\ga^{k'}(a_3 z) \left[a_3 z, a_0 z \right]_{k',k}
\ket{p(P,\la)}
\eea
%
\end{widetext}
%
where $\ket{p(P,\la)}$ denotes the proton state with 
momentum $P$, $P^2 = M^2$ and helicity $\la$. 
$u,d$ are the quark-field operators. The Greek letters $\al,\be,\ga$ 
stand for Dirac indices, the Latin letters $i,j,k$ refer to color.
$z$ is an arbitrary light-like  vector, $z^2=0$, the $a_i$ are real numbers.
The gauge-factors $\left[x,y\right]$ are defined as
\beq{path}
\left[x,y\right] = {\rm P}\exp\left[ig\int_0^1\! \dd t\, (x-y)_\mu 
A^\mu(t x + (1-t) y)\right]\,
\eeq
and render the matrix element in (\ref{dreiquark}) gauge-invariant.
To simplify the notation we will not write the gauge-factors explicitly
in what follows but imply that they are always present.

Taking into account Lorentz covariance, spin and parity of the
nucleon, the most general decomposition  of the matrix element in
\Gl{dreiquark} involves 24 invariant functions \cite{BFMS00}. To the 
leading twist-three accuracy only three of them are relevant. 
We can write in a shorthand notation
\bea{lt1}
&&4\bra{0} \ep^{ijk} u_\al^i(a_1 z) u_\be^j(a_2 z) d_\ga^k(a_3 z) 
\ket{p(P,\la)}_{\rm twist-3}
=
\nonumber\\&&{}= 
V^p_1 \,(v_1)_{\al\be,\ga} +  
A^p_1 \,(a_1)_{\al\be,\ga} +  
T^p_1 \,(t_1)_{\al\be,\ga}   
\eea
where 
\bea{Lor1}
 (v_1)_{\al\be,\ga}  &=&  \left(\!\not\!{p}C \right)_{\al \be} \left(\ga_5 N^+\right)_\ga 
\nonumber\\  
 (a_1)_{\al\be,\ga}  &=&  \left(\!\not\!{p}\ga_5 C \right)_{\al \be} N^+_\ga
\nonumber\\  
 (t_1)_{\al\be,\ga}  &=&  \left(i \si_{\perp p} C\right)_{\al \be} 
                            \left(\ga^\perp\ga_5 N^+\right)_\ga   
\eea
stand for the Lorentz structures.

In turn, $V^p_1$, $A^p_1$, $T^p_1$ can be written as
\beq{fourier}
F(a_i p\cdot z) = \int \! {\cal D} x\, e^{-ipz 
\sum_i x_i a_i} F(x_i)\,,
\eeq
where the functions $F(x_i)$ depend on the dimensionless
variables $x_i,\, 0 < x_i < 1, \sum_i x_i = 1$ which 
correspond to the longitudinal momentum fractions 
carried by the quarks inside the nucleon.  
The integration measure is defined as 
\beq{integration}
\int\! {\cal D} x  = \int_0^1\! \dd x_1 \dd x_2 \dd x_3\, 
\de (x_1 + x_2 + x_3 - 1)\,.
\eeq 
Note that 
\bea{sym1}
 (v_1)_{\al\be,\ga}  &=& (v_1)_{\be\al,\ga}\,,
\nonumber\\  
 (a_1)_{\al\be,\ga}  &=& -(a_1)_{\be\al,\ga}\,,
\nonumber\\
 (t_1)_{\al\be,\ga}  &=& (t_1)_{\be\al,\ga}\,.
\eea
Since the operator on the l.h.s. of \Gl{lt1} is symmetric under the 
exchange of the two $u$-quarks, this property implies that the 
$V$- and $T$-functions are symmetric and the $A$-function is
antisymmetric under the exchange of the first two arguments, respectively:
\bea{sym2}
  V^p(1,2,3) &=& V^p(2,1,3)\,,
\nonumber\\ 
  A^p(1,2,3) &=& -A^p(2,1,3)\,,
\nonumber\\
  T^p(1,2,3) &=&  T^p(2,1,3)\,.  
\eea
In addition, the matrix element in \Gl{lt1} has to fulfill 
the symmetry relation  
\bea{sss}
&&\bra{0} \ep^{ijk} u_\al^i(1) u_\be^j(2) d_\ga^k(3) \ket{P}
\nonumber\\
&&{}+\bra{0} \ep^{ijk} u_\al^i(1) u_\ga^j(3) d_\be^k(2 ) \ket{P}
\nonumber\\
&&{}+\bra{0} \ep^{ijk} u_\ga^i(3) u_\be^j(2) d_\al^k(1) \ket{P} = 0
\label{isospin}
\eea
that follows from the condition that the nucleon state has isospin 1/2:
\beq{operator}
\left(T^2 - \frac{1}{2}(\frac{1}{2} +1)\right)
\bra{0} \ep^{ijk} u_\al^i(1) u_\be^j(2) d_\ga^k(3) 
\ket{P} = 0\,,
\eeq 
where 
\beq{op2}
T^2 = \frac{1}{2} \left(T_+T_- + T_- T_+ \right) + T_3^2 \,
\eeq
and $T_\pm$ are the usual isospin step-up and step-down operators. 
Applying the set of Fierz transformations
\bea{f-t3}
(v_1)_{\ga\be,\al} &=& \frac{1}{2}\left(  v_1 -  a_1 -   t_1
\right)_{\al\be, \ga} \nn \\
(a_1)_{\ga\be,\al} &=& \frac{1}{2} \left(-   v_1 +   a_1 -   t_1
\right)_{\al\be, \ga} \nn \\
(t_1)_{\ga\be,\al} &=& - \left(v_1 +  a_1 \right)_{\al\be,\ga}
\eea
one ends up with the condition
\beq{isospin2}
2 T_1^p(1,2,3) = [V^p_1-A^p_1](1,3,2) + [V^p_1-A^p_1](2,3,1)\,, 
\eeq
which allows to express the tensor DA of the leading   
twist in terms of the vector and axial vector distributions. Since the 
latter have different symmetry, they can be combined together to define 
the single independent leading twist-3 proton DA
\beq{twist-3}
\Phi^p_3(x_1,x_2,x_3) = [V^p_1 - A^p_1](x_1,x_2,x_3) 
\eeq
which is well known and received a lot of attention in the literature.  
The neutron leading twist DA $\Phi^n_3(x_1,x_2,x_3)$ can readily be obtained
by the interchange of $u$ and $d$ quarks in the defining \Gl{lt1}. 
For all invariant functions $F=V,A,T$ proton and neutron DAs differ by an overall
sign:
\beq{proneu}
    F^p(1,2,3) = - F^n(1,2,3)\,, 
\eeq
as follows from the isospin symmetry. This property is retained for all twists.

\subsubsection{Pion-Nucleon Generalized Distribution Amplitudes of twist-3}

Our aim is to describe the generalized distribution amplitudes (GDA) of a pion-nucleon
system with small invariant mass in the similar formalism. To this end we 
define for the $p\pi^0$-system:   
%
\begin{widetext}
%
\beq{lt2}
4\bra{0} \ep^{ijk} u_\al^i(a_1 z) u_\be^j(a_2 z) d_\ga^k(a_3 z) 
\ket{p(P,\la)\pi^0}_{\rm tw-3}=
(\gamma_5)_{\ga\de}\frac{-i}{f_\pi}\left[
V^{p\pi^0}_1 (v_1)_{\al\be,\de} +  
A^{p\pi^0}_1 (a_1)_{\al\be,\de} +  
T^{p\pi^0}_1 (t_1)_{\al\be,\de}\right],
\eeq
and, similar, for $n\pi^+$:  
\beq{lt3}
4\bra{0} \ep^{ijk} u_\al^i(a_1 z) u_\be^j(a_2 z) d_\ga^k(a_3 z) 
\ket{n(P,\la)\pi^+}_{\rm tw-3}=
(\gamma_5)_{\ga\de}\frac{-i}{f_\pi}\left[
V^{n\pi^+}_1\! (v_1)_{\al\be,\de} +  
A^{n\pi^+}_1\! (a_1)_{\al\be,\de} +  
T^{n\pi^+}_1\! (t_1)_{\al\be,\de}\right].
\eeq
An extra $\gamma_5$ is needed to conserve parity. Similar to the proton 
case, the symmetry of the two $u$-quarks implies that the GDAs $V$ and $T$ are symmetric,
and $A$ is antisymmetric to the exchange of the first two arguments, respectively.

As observed in \cite{PPS01}, the  GDAs can be calculated using current algebra for small invariant 
masses of the pion-nucleon system in the  soft pion limit. 
One obtains
\beq{softinit}
\langle0|O|\pi^a(k)N_f(P,\lambda)\rangle =
  - \frac{i}{f_\pi} \langle 0|[Q_5^a, O] |N_f(P,\lambda)\rangle
-\frac{ig_A}{4f_\pi (P\cdot k)}
 \sum_{\lambda',f'}\langle 0|O|N_{f'}(P,\lambda')\rangle \, \bar N (P,\lambda')\!\not\!k \gamma_5 \tau^a_{f'f} N(P,\lambda)
\eeq 
and similar for the pion-nucleon final state
\beq{softfinal}
\langle\pi^a(k)N_f(P,\lambda)|O^\dagger|0\rangle =
 -\frac{i}{f_\pi} \langle N_f(P,\lambda)|[Q_5^a, O^\dagger]|0\rangle
+\frac{ig_A}{4f_\pi (P\cdot k)}
 \sum_{\lambda',f'}\bar N(P,\lambda)\not\!k\gamma_5\tau^a_{ff'}N(P,\lambda')
   \langle N_{f'}(P,\lambda')|O^\dagger|0\rangle.  
\eeq
%
\end{widetext}
%
Here $f_\pi=93$~MeV is the pion decay constant 
defined as $\langle 0|\bar q \gamma_\mu\gamma_5 \frac12 \tau^a q|\pi^b(k)\rangle =i \delta^{a b} f_\pi k_\mu$
and $g_A\simeq 1.25$ is the axial charge 
of the nucleon. $O$ is the nonlocal three-quark operator and 
$Q^a_5$ is the operator of the axial charge
\beq{Q5}
       Q_5^a = \int d^3 x\, \bar q(x)\gamma_0\gamma_5\frac{\tau^a}{2} q(x)\,,\qquad 
   q = \left( 
       \begin{tabular}{l} u\\d\end{tabular}
       \right). 
\eeq
where the $\tau^a$ are the usual Pauli matrices.
The second term in \Gl{softinit}, \Gl{softfinal} corresponds to the pion bremsstrahlung from the
ingoing (outgoing) nucleon. This term corresponds to a 
$\pi N$ system in P-wave and can, in principle, be separated from the 
first (commutator) contribution that is S-wave by considering the angular distributions 
in the $\pi N$ system. Note also that the bremsstrahlung contribution vanishes at the threshold
of pion production $W_{\rm th} =M+m_\pi$ but becomes significant for  $W-W_{\rm th} \sim m_\pi$. 
This contribution is determined in terms of the nucleon DAs that eventually combine to produce the 
nucleon electromagnetic form factors; it can always be 
added. In what follows we will concentrate on the S-wave (commutator) term.    

For our purposes we need to specify $O$ to be the relevant three-local light-ray operator. 
\beq{def:O}
O\to O^{uud}_{\al\be\ga} = \ep^{ijk} u_\al^i(a_1 z) u_\be^j(a_2 z) d_\ga^k(a_3 z)
\eeq
Here the superscript $(uud)$ indicates the flavor content and simultaneously 
 the order in which the quark fields are positioned on the light ray. For example, 
$O^{uud}_{\al\be\ga}$ and $O^{udu}_{\al\be\ga}$ correspond to the 
$u_\al(a_1z)u_\be(a_2z)d_\ga(a_3z)$ and $u_\al(a_1z)d_\be(a_2z)u_\ga(a_3z)$ configurations, respectively.
One obtains
%
\begin{widetext}
%
\bea{softin}
  \langle 0|O^{uud}_{\al\be\ga}|n(P,\lambda)\pi^+(k)\rangle &=& 
 \frac{1}{\sqrt{2}}\Big\{\langle 0|O_{\al\be\ga}^{uud}|n(P,\lambda)\pi^1(k)\rangle
  +i \langle 0|O_{\al\be\ga}^{uud}|n(P,\lambda)\pi^2(k)\rangle\Big\}  
\nonumber\\ 
 &=&- \frac{i}{\sqrt{2}f_\pi}
  \Big\{ \langle 0|[Q^1_5,O_{\al\be\ga}^{uud}]|n(P,\lambda)\rangle 
 +i\langle 0|[Q^2_5,O^{uud}_{\alpha\beta\gamma}]|n(P,\lambda)\rangle \Big\}+\ldots
\nonumber\\
 \langle 0|O^{uud}_{\alpha\beta\gamma}|p(P,\lambda)\pi^0(k)\rangle &=& -\frac{i}{f_\pi}
   \langle 0|[Q^3_5,O^{uud}_{\alpha\beta\gamma}]|p(P,\lambda)\rangle +\ldots
\eea
where the ellipses stand for bremsstrahlung contributions. 
Calculation of the commutators $[Q^1_5,O_{\al\be\ga}^{uud}(y)]$ and 
$[Q^3_5,O^{uud}_{\alpha\beta\gamma}(y)]$
reduces to a chiral rotation
\beq{chiral}
   [Q_5^a,q_f] = -\Bigg(\frac{\tau^a}{2}\Bigg)_{ff'}\gamma_5 q_{f'}\,.
\eeq
For the three-quark operators one uses the chain rule $[A,BCD] = [A,B]CD + B[A,C]D + BC[A,D]$
to obtain:
\bea{chiral2}
{}  [Q^1_5,O_{\al\be\ga}^{uud}] &=&  
   -\frac{1}{2}\big\{(\gamma_5)_{\al \lambda}O^{ddu}_{\lambda\gamma\beta} + (\gamma_5)_{\be \lambda}O^{ddu}_{\lambda\gamma\alpha}+(\gamma_5)_{\gamma \lambda}O^{uuu}_{\alpha\beta \lambda}\big\}\,, 
\nonumber\\
{}  [Q^2_5,O_{\al\be\ga}^{uud}] &=& 
                        \phantom{-}\frac{i}{2}\big\{(\gamma_5)_{\alpha \lambda} O^{ddu}_{\lambda\gamma\beta}
                                     +(\gamma_5)_{\beta \lambda}O^{ddu}_{\lambda\gamma\alpha}
                                     -(\gamma_5)_{\gamma \lambda}O^{uuu}_{\alpha\beta \lambda}\big\}\,,
\nonumber\\
{} [Q^3_5,O^{uud}_{\alpha\beta\gamma}] &=&- \frac{1}{2}
                                  \big\{(\gamma_5)_{\alpha \lambda} O^{uud}_{\lambda\beta\gamma}
                                     +(\gamma_5)_{\beta \lambda} O^{uud}_{\alpha \lambda\gamma}-
                                     (\gamma_5)_{\gamma \lambda} O^{uud}_{\alpha\beta \lambda}\big\}\,.
\eea
Taking the nucleon-to-vacuum matrix element, Eqs.~(\ref{lt1}), (\ref{lt2}) and (\ref{lt3}), 
and using symmetry relations (\ref{sym1}) and Fierz transformations (\ref{f-t3})
one obtains after some algebra
\bea{result1}
 A_1^{n\pi^+}(1,2,3)&=&\frac{1}{\sqrt{2}}\Big\{\frac{1}{2}V_1^n(1,3,2)
         -\frac{1}{2}V_1^n(2,3,1)-\frac{1}{2}A_1^n(2,3,1)+
            \frac{1}{2}A_1^n(1,3,2) +T_1^n(2,3,1)-T_1^n(1,3,2)\Big\},
\nonumber\\
V_1^{n\pi^+}(1,2,3)&=&
\frac{1}{\sqrt{2}}\Big\{\frac{1}{2}V_1^n(1,3,2)+\frac{1}{2}A_1^n(1,3,2)+T_1^n(1,3,2)
+\frac{1}{2}V_1^n(2,3,1)+\frac{1}{2}A_1^n(2,3,1)+T_1^n(2,3,1)
\Big\}, 
\nonumber\\
T_1^{n\pi^+}(1,2,3)&=&
\frac{1}{2\sqrt{2}}\Big\{A_1^n(2,3,1)+A_1^n(1,3,2)
-V_1^n(2,3,1)-V_1^n(1,3,2)\Big\},
\eea
and
\bea{result2}
 {V}_1^{p\pi^0}(1,2,3) &=&\frac{1}{2}V_1^p(1,2,3)\,,
 {}\nonumber\\
 {A}_1^{p\pi^0}(1,2,3) &=&\frac{1}{2}A_1^p(1,2,3)\,,
 \nonumber\\
{T}_1^{p\pi^0}(1,2,3) &=& \frac{3}{2}T_1^p(1,2,3)\, ,
\eea 
which is the desired result. 
One can further simplify (\ref{result1}) using the substitution in \Gl{isospin2} to obtain:
\bea{result3}
{V}_1^{n\pi^+}(1,2,3) &=&
 \frac{1}{\sqrt{2}} \Big\{V_1^n(1,3,2)+V_1^n(1,2,3)+V_1^n(2,3,1)+A_1^n(1,3,2)+A_1^n(2,3,1)\Big\},
\nonumber\\
{A}_1^{n\pi^+}(1,2,3) &=& 
 -\frac{1}{\sqrt{2}}       
            \Big\{V_1^n(3,2,1)-V_1^n(1,3,2)+A_1^n(2,1,3)+A_1^n(2,3,1)+A_1^n(3,1,2)\Big\},      
\nonumber\\
T_1^{n\pi^+}(1,2,3)&=&
\frac{1}{2\sqrt{2}}\Big\{A_1^n(2,3,1)+A_1^n(1,3,2)-V_1^n(2,3,1)-V_1^n(1,3,2)\Big\}.
\end{eqnarray}
%
\end{widetext}
%
Note that the functions $V,A,T$ for the both cases $n\pi^+$ and $p\pi^0$ satisfy the 
symmetry conditions (\ref{sym2}) which follow directly from their definitions in (\ref{lt2}),
(\ref{lt2}), respectively. On the other hand, the isospin relation similar to 
(\ref{isospin2}) is not valid, since the pion-nucleon pair can have both
isospin 1/2 and 3/2.

\subsection{Higher  Twists } 
Taking into account Lorentz covariance, spin and parity of the nucleon and the pion, the most general 
decomposition of the three-quark matrix element in Eq.~(\ref{dreiquark}) involves 24 invariant 
functions. In total, there are three leading-twist-3 invariant functions ($V_1,A_1,T_1$), nine of twist-4
($V_2,A_2,V_3,A_3,S_1,P_1,T_2,T_3,T_7$), nine of twist-5 ($V_4,A_4,V_5,A_5,S_2,P_2,T_4,T_5,T_8$),
and three of twist-6 ($V_6,A_6,T_6$). 
Using the shorthand notation of Ref.~\cite{BFMS00} for the relevant Lorentz structures, we define
%
\begin{widetext}
%
\begin{eqnarray}
 &&\hspace{-3.475cm}4\cdot\bra{0} \ep^{ijk} u_\al^i(a_1z) u_\be^j(a_2z) d_\ga^k(a_3z) 
\ket{N(P,\la)\pi(k)}=
\nonumber\\
 =(\gamma_5)_{\ga\de}\frac{-i}{f_{\pi}}&\Bigg[&
S_1^{\pi N}(s_1)_{\alpha\beta,\de}+S_2^{\pi N}(s_2)_{\alpha\beta,\de}+
P^{\pi N}_1 (p_1)_{\al\be,\de}+P_2^{\pi N}(p_2)_{\alpha\beta,\de} \hspace{0.6cm}
\nonumber\\
&+&V_1^{\pi N}(v_1)_{\alpha\beta,\de}+V_2^{\pi N}(v_2)_{\alpha\beta,\de}+
\frac{1}{2}V_3^{\pi N}(v_3)_{\alpha\beta,\de}+\frac{1}{2}V_4^{\pi N}(v_4)_{\alpha\beta,\de}
\nonumber\\&+&V_5^{\pi N}(v_5)_{\alpha\beta,\de}+V_6^{\pi N}(v_6)_{\alpha\beta,\de}
+A_1^{\pi N}(a_1)_{\alpha\beta,\de}
+A_2^{\pi N}(a_2)_{\alpha\beta,\de}
\nonumber\\
&+&\frac{1}{2}A_3^{\pi N}(a_3)_{\alpha\beta,\de}+\frac{1}{2}A_4^{\pi N}(a_4)_{\alpha\beta,\de}
+A_5^{\pi N}(a_5)_{\alpha\beta,\de}
+A_6^{\pi N}(a_6)_{\alpha\beta,\de}
\nonumber\\
&+&T_1^{\pi N}(t_1)_{\alpha\beta,\de}+T_2^{\pi N}(t_2)_{\alpha\beta,\de}
+T_3^{\pi N}(t_3)_{\alpha\beta,\de}+T_4^{\pi N}(t_4)_{\alpha\beta,\de}
\nonumber\\
&+&T_5^{\pi N}(t_5)_{\alpha\beta,\de}+T_6^{\pi N}(t_6)_{\alpha\beta,\de}
+\frac{1}{2}T_7^{\pi N}(t_7)_{\alpha\beta,\de}+\frac{1}{2}T_8^{\pi N}(t_8)_{\alpha\beta,\de}~~\Bigg]~.
\label{npi+}
\end{eqnarray}
%
\end{widetext}
%
The calculation of higher-twist pion-nucleon GDAs is similar to the leading twist, but 
proves to be much more cumbersome. The complete set of the necessary Fierz identities 
for the Lorentz structures $(v_i)_{\alpha\beta,\de}$, $(a_i)_{\alpha\beta,\de}$ $(t_i)_{\alpha\beta,\de}$
is given in  Ref.~\cite{BFMS00}. The results for the GDAs of all twists are collected 
in Appendix~A below. In addition, in Appendix~B we present the results for the functions ${\mathcal V}_1^M$ and
${\mathcal A}_1^M$ that appear in the off-light-cone $\mathcal{O}(x^2)$ corrections in the OPE 
involving three-quark currents, 
\bea{lt33}
&&
4\bra{0} \ep^{ijk} u_\al^i(a_1 x) u_\be^j(a_2 x) d_\ga^k(a_3 x) 
\ket{N\pi} =
\nonumber\\ 
&&=
(\gamma_5)_{\ga\de}\frac{-i}{f_\pi}\Big\{
\left[V_1 +\frac{x^2m_N^2}{4}{\mathcal V}_1^M\right] (v_1)_{\al\be,\de}
\nonumber\\&&{} +  
\left[A_1 +\frac{x^2m_N^2}{4}{\mathcal A}_1^M\right]  (a_1)_{\al\be,\de}
\nonumber\\&&{} +  
\left[T_1 +\frac{x^2m_N^2}{4}{\mathcal T}_1^M\right] (t_1)_{\al\be,\de}\Big\}^{\pi N}+\ldots,
\eea
for the case that the light-cone positions of two of the three quarks coincide,  
to the first subleading order in the conformal expansion.  The function ${\mathcal T}_1^M$
does not contribute to the sum rules to our accuracy.


\section{Light Cone Sum Rules}
\setcounter{equation}{0}
\subsection{The soft pion limit}
{}For technical reasons, it is convenient to write the  sum rules for the complex conjugated amplitude
with the pion-nucleon pair in the initial state. To this end we consider the correlation function
\bea{correlator} 
\lefteqn{T_{\nu}^{\pi N}(P,q) =}
\\&=& i\! \int\! \dd^4 x \, e^{i q x}  
\bra{0} T\left\{\eta_p (0) j_{\nu}^{\mathrm{em}}(x)\right\} \ket{N(P) \pi(k)},
\nonumber
\eea
where 
\bea{current}
&&\eta_p(x) = \ep^{ijk} \left[u^i(x) C\ga_\mu u^j(x)\right]\,\ga_5 \ga^\mu d^k(x)\,,
\nonumber\\&&{}\bra{0} \eta_p(0)\ket{N(P)}  = \lambda_1^p m_N N(P) \,,
\eea
is the so-called Ioffe proton interpolating current \cite{Ioffe:1981kw}, $\lambda_1^p$ is  the corresponding coupling.

In the limit $|k|\to 0$ for fixed $q^2$ and $(P')^2 = (P+k-q)^2$ the correlation function in (\ref{correlator})
can be calculated using PCAC and current algebra in terms of the correlation functions without a pion and
involving chirally-rotated currents 
\begin{eqnarray}
\lefteqn{  T_\nu^{\pi N} (P,q) =}
\nonumber\\
&=& -\frac{i}{f_\pi}\left[
 i \int d^4x\, e^{iqx}\langle 0|T\{{[Q^a_5,\eta_p(0)]}j_\nu^{\mathrm{em}}(x)\}|N(P)\rangle\right.
\nonumber\\
&&{}\left.\hspace*{0.65cm}
+i \int d^4x\, e^{iqx}\langle 0|T\{\eta_p(0){[Q^a_5,j_\nu^{\mathrm{em}}(x)]\}}|N(P)\rangle
   \right]
\nonumber\\
&&{}
+\mbox{\rm{bremsstrahlung~terms}}\,,
\label{LET1}
\end{eqnarray} 
where $Q^a_5$ is the axial charge and the bremsstrahlung contributions correspond to pion absorption
by the initial state nucleon. 

The commutators can easily be evaluated with the result
\bea{commu1}
    {}[Q^+_5,j_\nu^{\mathrm{em}}(x)]  &=& -\frac{1}{\sqrt{2}} A_\nu(x) \,, 
 \quad A_\nu = \frac{1}{\sqrt{2}}\bar q \gamma_\nu\gamma_5\tau^+q\,,
\nonumber\\
    {}[Q^3_5,j_\nu^{\mathrm{em}}(x)]  &=& 0
\eea
where 
$\tau^\pm = \frac{1}{\sqrt{2}}(\tau^1\pm i\tau^2)$, $Q_5^\pm = \frac{1}{\sqrt{2}}(Q_5^1\pm iQ_5^2)$ etc.
Also
\bea{commu2}
   {}[Q^+_5,\eta_p(x)] &=& -\frac{1}{\sqrt{2}} \gamma_5 \eta_n(x)\,,
\nonumber\\
   {}[Q^3_5,\eta_p(x)] &=& -\frac12 \gamma_5 \eta_p(x)\,,
\eea
where $\eta_n$ is the neutron current
\beq{currentn}
\eta_n(x) = -\ep^{ijk} \left[d^i(x) C\ga_\mu d^j(x)\right]\,\ga_5 \ga^\mu u^k(x)\,.
\eeq
\begin{figure}[ht]
\centerline{\epsfxsize7cm\epsffile{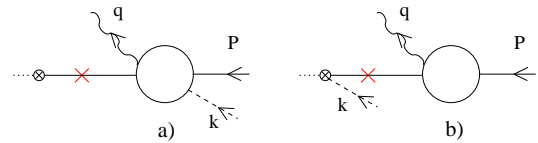}}
\caption{\small
Schematic structure of the pole terms in the correlation function
 (\ref{correlator})
}
\label{figdisp}
\end{figure}
The contributions of interest to (\ref{correlator}) are those singular in the vicinity of $P'^2 \to m_N^2$,
see Fig.~\ref{figdisp}. Note that in addition to the nucleon pole, Fig.~\ref{figdisp}a, one has to take into account the 
semidisconnected contribution with the pion-nucleon intermediate state, Fig.~\ref{figdisp}b.
{}For example, for $n\pi^+$ one obtains 
\begin{widetext}
\begin{eqnarray}
T_{\nu}^{\pi^+ n}(P,q)
&=& \frac{i\lambda_1^p m_N}{f_\pi} \frac{m_N+\not\!P'}{m_N^2-P'^2}\gamma_5
 \left\{(\gamma_\nu q^2 -q_\nu \not\!q) \frac{G_1^{n\pi^+}}{m_N^2}-
\frac{i\sigma_{\nu\mu}q^\mu}{2m_N}G_2^{n\pi^+}\right\}N(P)
\nonumber\\
&&{}- \frac{i\lambda_{n\pi^+}m_N}{f_\pi} \gamma_5 \frac{m_N+\not\!P'-\not\!k}{m_N^2-(P'-k)^2}
\left\{\gamma_\nu F_1^{n}-
\frac{i\sigma_{\nu\mu} q^\mu}{2m_N}F_2^{n}\right\}N(P)+\ldots
\label{rep1}
\end{eqnarray}
where the ellipses stand for less singular contributions, $F_1^n$ and 
$F_2^n$ are Dirac and Pauli electromagnetic neutron form factors, respectively
\begin{eqnarray}
\langle N(P')|j_{\mu}^{\rm em}(0) |N(P)\rangle =
\bar{N}(P')\left[\gamma_{\mu}{F_1^N(Q^2)}-i\frac{\sigma_{\mu\nu}q^{\nu}}{2m_N}{ F_2^N(Q^2)}\right]N(P)
\end{eqnarray} 
and $\lambda_{n\pi^+}$ is the coupling of the Ioffe current to the $n\pi^+$ state:
\beq{Ioffe-piN}
\bra{0} \eta_p(0)\ket{N(P'-k)\pi(k)}  = -\frac{i}{f_\pi} \lambda_{\pi N} m_N \gamma_5 N(P'-k) \,.
\eeq
On the other hand, using the representation in Eq.~(\ref{LET1}) instead, one obtains
\begin{eqnarray}
T_{\nu}^{\pi^+ n}(P,q)
&=&  \frac{i\lambda_1^p m_N}{f_\pi}\frac{1}{\sqrt{2}} \frac{m_N+\not\!P'}{m_N^2-P'^2}
 \left\{
\gamma_\nu {G_A(Q^2)} - \frac{q_\nu}{2 m_N}  {G_P(Q^2)}
-  i \frac{\si_{\nu\mu}  q^\mu}{2 m_N}  {G_T(Q^2)} 
\right\}\gamma_5N(P)
\nonumber\\
&&{} +\frac{i}{\sqrt{2}f_\pi}\gamma_5\lambda_1^{n}m_N \frac{m_N+\not\!P'}{m_N^2-P'^2}
\left\{\gamma_\nu F_1^{n}-
\frac{i\sigma_{\nu\mu} q^\mu}{2m_N}F_2^{n}\right\}N(P)+\ldots
\label{rep2}
\end{eqnarray}
where the form factors in the first line are defined as
\begin{eqnarray}
\langle N(P')|A_{\nu}(0) |N(P)\rangle =
\bar{N}(P') \Big[ \gamma_\nu {G_A(Q^2)} - \frac{q_\nu}{2 m_N}  {G_P(Q^2)}
-  i \frac{\si_{\nu\mu}  q^\mu}{2 m_N}  {G_T(Q^2)}  \Big] \gamma_5 N(P)\,.
\end{eqnarray}
\end{widetext}
Note that $G_T(Q^2)=0$ because of the isospin and CP invariance. 

In the same approximation the pion-nucleon coupling is given by $\lambda_{n\pi^+} = 
 \left(-\frac{1}{\sqrt{2}}\right)\lambda_1^p$
(cf. Eq.~(\ref{commu2})) so that the terms in the second line in Eqs.~(\ref{rep1}) and (\ref{rep2}) coincide:
The contribution of the chiral rotation of the nucleon current is identically equal to the semidisconnected 
contribution in Fig.~\ref{figdisp}b of the pion coupling to the current. Equating the remaining contributions
in the first line of Eqs.~(\ref{rep1}) and (\ref{rep2}) one obtains
\bea{softpi1}
  Q^2 G_1^{n\pi^+}(Q^2) &=& \frac{m_N^2}{\sqrt{2}}G_A(Q^2) + \mathcal{O}(m_\pi/\Lambda, |k|/\Lambda)\,,   
\nonumber\\
  G_2^{n\pi^+}(Q^2) &=& 0  + \mathcal{O}(m_\pi/\Lambda, |k|/\Lambda)\,, 
\eea   
which is the classical result \cite{Kroll:1953vq,Nambu:1997wb,Vainshtein:1972ih,Scherer:1991cy}.
The consideration  of the $p\pi^0$ state is similar, the only difference being that the commutator of the 
electromagnetic current with the axial charge vanishes in this case, so that the both amplitudes vanish 
at threshold, up to corrections in $m_\pi$ and/or $|k|$:  
\bea{softpi2}
  Q^2 G_1^{p\pi^0}(Q^2) &=& 0 + \mathcal{O}(m_\pi/\Lambda, |k|/\Lambda)\,,   
\nonumber\\
  G_2^{p\pi^0}(Q^2) &=& 0  + \mathcal{O}(m_\pi/\Lambda, |k|/\Lambda)\,. 
\eea   
To avoid misunderstanding, note that our (schematic) derivation does not include 
bremsstrahlung contributions, which have to be added.   
The derivation breaks down for large momentum transfers, however. One way to see this is that 
the limit $k \to 0$ that is implied in the current algebra relations, becomes far away from the physical
region. In particular, the position of the pole in the pion-nucleon intermediate state in Fig.~\ref{figdisp}b
$(P'-k)^2 = m_N^2$ moves away from the nucleon pole at $(P')^2 = m_N^2$. Requiring that 
$(P'-k)^2 - (P')^2 \ll \Lambda^2_{\mathrm QCD} \sim m^2_N$ one obtains a restriction 
for the applicability of the soft pion limit $Q^2 \ll \Lambda^3/m_\pi$. 
In what follows we develop a technique to calculate the amplitudes in the opposite limit 
of large momentum transfers, $Q^2 \gg \Lambda^3/m_\pi$.
      
\subsection{The sum rules}
The general philosophy of the LCSR approach is explained in the Introduction.
The main idea is to separate contributions of small distances to the correlation 
function in (\ref{correlator}) in terms of calculable coefficient functions in 
front of operator matrix elements and apply the soft pion techniques for the 
evaluation of the latter rather than the correlation function itself. 
Information about the generalized form factors can then be extracted by matching
the QCD calculation at moderate negative values $P'^2 \sim - 1$~GeV$^2$ with the 
dispersion representation (\ref{rep1}) in terms of hadronic states.  

The correlation function in (\ref{correlator}) involves several invariant functions
that can be separated by the appropriate Lorentz projections. The structures that are 
most useful for writing the LCSRs are usually those containing the maximum 
power of the large momentum $p^+ \sim pz$. We define
\beq{project4}
  \Lambda_+ T_z = + \frac{i}{f_\pi}(pz+kz)\gamma_5 \left\{ m_N \mathcal{A} + 
    \!\not\!q_\perp \mathcal{B}\right\} N^+(P)\,.
\eeq
The invariant functions $\mathcal{A}$ and  $\mathcal{B}$ can be calculated for sufficiently 
large Euclidean $Q^2$ and $P'^2=(P+k-q)^2$ in terms of the generalized distribution amplitudes of the 
pion-nucleon system using the operator product expansion. 
We use the following notations:
\bea{notat1}
\widetilde{F}(x_3)
 &=& \int_1^{x_3}\!\!dx'_3\int_0^{1- x^{'}_{3}}\!\!\!dx_1\, F(x_1,1-x_1-x'_3,x'_3)\,,
\nn\\
\lefteqn{\hspace*{-0.9cm}\widetilde{\!\widetilde{F}}(x_3)= \int_1^{x_3}\!\!dx'_3 \int_1^{x'_3}\!\!dx^{''}_3  }
\nn\\[-2mm]&&\hspace*{-0.2cm}\times 
\int_0^{1- x^{''}_{3}}\!\!\!\!dx_1\, F(x_1,1-x_1-x^{''}_3,x^{''}_3)
\eea
and
\bea{notat2}
\widehat{F}(x_2) 
&=& \int_1^{x_2}\!\!dx'_2\int_0^{1-x^{'}_{2}}\!\!\!dx_1 F(x_1,x'_2,1-x_1-x'_2)\,,
\nn\\
\lefteqn{\hspace*{-0.9cm}\widehat{\!\widehat{F}}(x_2)= \int_1^{x_2}\!\!dx'_2 \int_1^{x'_2}\!\!dx^{''}_2}
\nn\\[-2mm] &&{}\hspace*{-0.2cm} \times
\int_0^{1-x^{''}_{2}}\!\!\!dx_1 F(x_1,x^{''}_2,1-x_1-x^{''})\,,
\eea
where $F=A,V,T$ is a generic pion-nucleon GDA that depends on the three valence quark momentum fractions.

To the three-level accuracy $\mathcal{A}^{\pi N}$  and $\mathcal{B}^{\pi N}$ are  given by the same expressions \cite{Braun:2006hz}
as the LCSRs for the proton electromagnetic form factors $F_1$ and $F_2$, respectively, with the substitution 
of the nucleon DAs by the pion-nucleon ones:
%
\begin{widetext}
%
\bea{Ioffe-em}
  \mathcal{A}^{\pi N} &= & \phantom{-}2e_d\int_0^1 dx_3\left\{
     \frac{Q^2+q_3^2}{q_3^4} \widetilde{V}_{123}^{\pi N}+\frac{x_3}{q_3^2}\int_0^{\bar x_3} dx_1 V_3^{\pi N}(x_i)+
      \frac{x_3^2m_N^2}{q_3^4}\widetilde{V}_{43}^{\pi N}\right\}
\nn\\
&&{} + 2 e_u \int_0^1 dx_2 \left\{
      \frac{x_2}{q_2^2} \int_0^{\bar x_2}dx_1 \left[ -2 V_1^{\pi N} + 3V_3^{\pi N}+A_3^{\pi N} \right](x_i)
       - \frac{2x_2 m_N^2}{q_2^4}\mathcal{V}_1^{\pi N,M(u)} +\frac{Q^2-q_2^2}{q_2^4}\widehat{V}_{123}^{\pi N}
\right.
\nn\\
&&{}\left. \hspace{0.0cm}
       + \frac{Q^2+q_2^2}{q_2^4}\widehat{A}_{123}^{\pi N}
       -\frac{x_2^2 m_N^2}{q_2^4}\left[\widehat{V}_{1345}^{\pi N} - 2\widehat{V}_{43}^{\pi N}+ \widehat{A}_{34}^{\pi N}\right] 
      -\frac{2x_2 m_N^2}{q_2^4}\widehat{\widehat{V}}_{123456}^{\pi N}
\right\},
\nn\\
 \mathcal{B}^{\pi N} &=& -2e_d\int_0^1 dx_3\left\{
    \frac{1}{q_3^2}\int_0^{\bar x_3} dx_1 V_1^{\pi N}(x_i) +\frac{m_N^2}{q_3^4}\mathcal{V}_1^{\pi N,M(d)}
     - \frac{x_3 m_N^2}{q_3^4} \left[\widetilde{V}_{123}^{\pi N} -  \widetilde{V}_{43}^{\pi N} \right]
     \right\}
\nn\\ &&{}
 + 2 e_u \int_0^1 dx_2 \left\{
    \frac{1}{q_2^2}\int_0^{\bar x_2}dx_1 \left[V_1+A_1\right]^{\pi N}(x_i)
      +\frac{m_N^2}{q_2^4}\left[\mathcal{V}_1^{\pi N,M(u)}+\mathcal{A}_1^{\pi N,M(u)}\right]
        \right.
\nn\\ &&{}\left. 
\hspace{0.0cm}
      +\frac{x_2m_N^2}{q_2^4} \left[ 
\widehat{V}_{1345}^{\pi N} +  \widehat{V}_{123}^{\pi N}+\widehat{A}_{123}^{\pi N} - 2\widehat{V}_{43}^{\pi N}+\widehat{A}_{34}^{\pi N}\right]
\right\},
\eea
%
\end{widetext}
%
where we used shorthand notations for the combinations of the DAs:
\bea{notat31}
&&\hspace*{-3mm}V_{43} = V_4-V_3\,,
\nn\\
&&\hspace*{-3mm}V_{123} = V_1-V_2-V_3\,,
\nn\\
&&\hspace*{-3mm}V_{1345} = -2V_1+V_3+V_4+2 V_5\,,
\nn\\
&&\hspace*{-3mm} V_{12345} = 2 V_1-V_2-V_3-V_4-V_5\,,
\nn\\
&&\hspace*{-3mm}  V_{123456} = -V_1+V_2+V_3+V_4+V_5-V_6\,,
\eea
\bea{notat32}
&&\hspace*{-3mm}A_{34} = A_3-A_4\,,
\nn\\
&&\hspace*{-3mm}A_{123} = -A_1+A_2-A_3\,,
\nn\\
&&\hspace*{-3mm} A_{1345} = -2A_1-A_3-A_4+2 A_5\,,
\nn\\
&&\hspace*{-3mm} A_{12345} = 2 A_1-A_2+A_3+A_4-A_5\,,
\nn\\
&&\hspace*{-3mm}  A_{123456} = A_1-A_2+A_3+A_4-A_5+A_6
\eea
and also 
\bea{notat33}
&&\hspace*{-3mm}T_{137}   = T_1 - T_3 - T_7\,,
\nn\\
&&\hspace*{-3mm}T_{13478}  = 2T_1 -T_3 - T_4 - T_7 - T_8\,,
\nn\\
&&\hspace*{-3mm}T_{134678} = T_1 -T_3 - T_4 + T_6 - T_7 - T_8 \,.
\eea
Explicit expressions for the functions $\mathcal{V}_1^{\pi N,M(u,d)}$ and $\mathcal{A}_1^{\pi N,M(u,d)}$
are given in Appendix~B. The superscripts $(u)$ or $(d)$ stand for the cases that one of the $u$-quarks
has the same position as the $d$-quark, and the two $u$-quarks have the same position, respectively.
In addition, we use a compact notation
\beq{q3q2}
q_3 \equiv q-x_3 (P+k)\,,\qquad q_2 \equiv q-x_2 (P+k)\,.   
\eeq
In the both  expressions in (\ref{Ioffe-em}) the functions with a ``tilde'' and a ``hat'' have $x_3$ and $x_2$ as
an argument, respectively, cf. (\ref{notat1}), (\ref{notat2}).
Also, in the terms involving two integrations over the momentum fractions, the remaining
momentum fraction is replaced by using $x_1+x_2+x_3=1$.

The Borel transformation and the continuum subtraction  
are performed by using the following substitution rules:  
%
\begin{widetext}
%
\bea{borel0}  
\int \dd x \frac{\varrho(x) }{(q-x(P+k))^2} &=&  
- \int_0^1 \frac{\dd x }{x} \frac{\varrho(x)}{(s - {P'}^2)}  
  \,\to\,  
-   \int_{x_0}^1 \frac{\dd x}{x} \varrho(x)  
\exp{ \left( - \frac{\bar x Q^2}{x M^2}  - \frac{\bar x m_N^2}{M^2}\right)}\, ,    
\\
\int \dd x \frac{\varrho(x) }{(q-x(P+k))^4} &=&  
\int_0^1 \frac{\dd x }{x^2} \frac{\varrho(x)}{(s - {P'}^2)^2}  
\,\to\,  
\frac{1}{M^2} \int_{x_0}^1 \frac{\dd x}{x^2} \varrho(x)  
\exp{  
\left( - \frac{\bar x Q^2}{x M^2}  - \frac{\bar x m_N^2}{M^2}\right)}    
+  
\frac{\varrho(x_0)\,e^{-s_0 /M^2} }{Q^2 + x_0^2 m_N^2} \, ,    
\nonumber
\eea  
where $M$ is the Borel parameter, 
$s = \frac{1-x}{x} Q^2  + (1-x) m_N^2$ and 
$x_0$ is the solution of the corresponding quadratic equation for $s = s_0$:
\bea{x0}
   x_0 &=&\bigg[ \sqrt{(Q^2+s_0-m_N^2)^2+ 4 m_N^2 Q^2}-(Q^2+s_0-m_N^2)\bigg]
   /(2m_N^2)\,.
\eea  
%
\end{widetext}
%
The contribution $\sim e^{-s_0 /M^2}$ in \Gl{borel0}
corresponds to the ``surface terms'' arising from  the  partial integration 
to reduce the power in the denominator $(q - x P)^{4} = (s  - {P'}^2 )^{4} (-x)^{4}$ 
to the usual dispersion representation with the denominator  
$\sim (s - {P'}^2 )$. Without continuum subtraction, i.e. in the 
limit  $s_0 \to \infty$ this term vanishes.

On the other hand, the nucleon contribution of interest corresponds to a pole 
term in the variable $P'^2$ at $P'^2=m_N^2$. For the relevant projections we get
\bea{projections-lhs}
 \mathcal{A}^{\pi N} & = & \phantom{-} \frac{2\lambda_1^p (Q^2/m_N^2) G^{\pi N}_1}{m_N^2- P'^2} + 
              \frac{2\lambda_{\pi N}F_1(Q^2)}{m_N^2-(P'-k)^2},
\nonumber\\  
 \mathcal{B}^{\pi N} & = &  {-}\frac{\lambda_1^p G^{\pi N}_2}{m_N^2- P'^2} +
   \frac{\lambda_{\pi N}F_2(Q^2)}{m_N^2-(P'-k)^2}
\eea
where $F_1$ and $F_2$ are the Dirac and Pauli proton (for $p\pi^0$) and neutron 
(for $n\pi^+$) electromagnetic 
form factors, respectively. The pion-nucleon coupling $\lambda_{\pi N} \equiv \lambda_{\pi N}(P'^2)$ is defined 
in Eq.~(\ref{Ioffe-piN}) and in general depends on the invariant mass $P'^2$ of the pion-nucleon system. 
Close to threshold in the $P'^2$-channel,  
$P'^2_{\mathrm{th}} = m_N^2(1+\delta)^2$, one obtains in the soft-pion limit 
$\lambda_{n\pi^+} = -\lambda_1^p/\sqrt{2}$,  $\lambda_{p\pi^0} = -\lambda_1^p/2$.

Note that the Borel transformation amounts in this case to the substitution
\bea{borel5}
 &&\frac{1}{m_N^2-P'^2} \to e^{-m_N^2/M^2}\,,
 \\
 &&\frac{1}{m_N^2-(P'-k)^2} \to (1+\delta) e^{-[m_N^2(1+\delta)^2 +\delta Q^2]/M^2}\,.
\nonumber
\eea
One observes that the contribution of the pion-nucleon intermediate state is suppressed
compared to the nucleon contribution of interest by an extra factor 
$\exp\{\delta[m_N^2(2+\delta)+Q^2]/M^2\}$. 
Requiring that this suppression is at least as strong as $\exp\{-[s_0-m_N^2]/M^2\}$ in which case
the pion-nucleon state has to be considered as a part of the continuum, we obtain
\bea{condition} 
&&\delta[m_N^2(2+\delta)+Q^2] \ge s_0-m_N^2\,,
\\ 
&&Q^2 \ge (s_0-m_N^2)/\delta -(2+\delta)m_N^2
\eea
where from $Q^2 \ge 7.3$~GeV$^2$. For smaller momentum transfers the pion-nucleon contribution 
has to be calculated explicitly and subtracted. The problem is that the corresponding couplings 
$\lambda_{\pi N}$ have to be taken at the invariant mass of the pion-nucleon pair $P'^2 = m_N^2(1+\delta)^2 +\delta Q^2$ 
and this dependence is not known {\it a priory}.   

%
\begin{widetext}
%
  
Neglecting the pion-nucleon intermediate state, we obtain the sum rules:
\bea{LCSRIoffe}
Q^2G^{\pi N}_1(Q^2)  & = & \frac{m_N^2}{2 \lambda_1^p}
\left[ \int_{x_0}^1 \dd x \left( - \frac{\varrho_2^a(x)}{x}  + 
\frac{\varrho_4^a(x)}{x^2M^2} \right)
        \exp{ \left( - \frac{\bar x Q^2}{x M^2}  + \frac{x 
m_N^2}{M^2}\right)}\,
+
\frac{\varrho_4^a(x_0)\,e^{-(s_0-m_N^2) /M^2} }{Q^2 + x_0^2 m_N^2} \right],
\nn
\\
G^{\pi N}_2(Q^2) & = & -\frac{1}{\lambda_1^p}
\left[ \int_{x_0}^1 \dd x \left( - \frac{\varrho_2^b(x)}{x}  + 
\frac{\varrho_4^b(x)}{x^2M^2} \right)
        \exp{ \left( - \frac{\bar x Q^2}{x M^2}  + \frac{x 
m_N^2}{M^2}\right)}\,
+
\frac{\varrho_4^b(x_0)\,e^{-(s_0-m_N^2) /M^2} }{Q^2 + x_0^2 m_N^2} \right],
\eea
where the functions $\varrho^{a,b}_{2,4}(x)$ are given in terms of the generalized pion-nucleon distribution 
amplitudes as
\bea{abbrev}
\varrho_2^{a}(x) & = &
       2e_d \bigg\{ \widetilde{V}_{123}^{\pi N} + x \int_0^{\bar x} dx_1 V_3^{\pi N}(x_i)
\bigg\}
  + 2 e_u \bigg\{ x \int_0^{\bar x}dx_1 \left[ -2 V_1 + 3V_3+A_3 
\right]^{\pi N}(x_i)
        - \widehat{V}_{123}^{\pi N} + \widehat{A}_{123}^{\pi N} \bigg\},
\nn \\
\varrho_4^{a}(x) & = &
       2e_d \bigg\{Q^2 \widetilde{V}_{123}^{\pi N}+ x^2 m_N^2 
\widetilde{V}_{43}^{\pi N}\bigg\}
+ 2 e_u  \bigg\{ Q^2 \left( \widehat{V}_{123}^{\pi N} + \widehat{A}_{123}^{\pi N} \right)
- x^2 m_N^2 \left[\widehat{V}_{1345}^{\pi N} - 2\widehat{V}_{43}^{\pi N}+ \widehat{A}_{34}^{\pi N}\right]
 \nn \\ && {}
- 2x m_N^2  \left( \mathcal{V}_1^{\pi N,M(u)} + \widehat{\widehat{V}}_{123456} ^{\pi N}\right) \bigg\},
\nn \\
\varrho_2^{b}(x) & = &
- 2 e_d \left\{ \int_0^{\bar x} dx_1 V_1^{\pi N}(x_i)    \right\}
+ 2 e_u \left\{ \int_0^{\bar x}dx_1 \left[V_1+A_1\right]^{\pi N}(x_i) \right\},
\nn \\
\varrho_4^{b}(x) & = &  -2e_d m_N^2
\bigg\{ \mathcal{V}_1^{\pi N,M(d)} - x \left[\widetilde{V}_{123} - 
\widetilde{V}_{43} \right]^{\pi N} \bigg\}
  + 2 e_u m_N^2 \bigg\{ \left[\mathcal{V}_1^{\pi N,M(u)}+\mathcal{A}_1^{\pi N,M(u)}\right]
\nn \\ &&{}
  + x \left[\widehat{V}_{1345} + 
\widehat{V}_{123}+\widehat{A}_{123}-2\widehat{V}_{43}+\widehat{A}_{34}\right]^{\pi N}
\bigg\}.
\eea
Note that the sum rules have exactly the same structure as the sum rules for electromagnetic nucleon form factors 
in Ref.~\cite{Braun:2006hz}, with the only difference that nucleon DAs are replaced by the pion-nucleon ones.

{}For completeness, we quote also the sum rule for the axial form factor of the proton from Ref.~\cite{Braun:2006hz}
(for the charged current defined in (\ref{commu1})):
\beq{GAIoffe}
G_A(Q^2)  =  \frac{1}{\lambda_1^p}
\left[ \int_{x_0}^1 \dd x \left( - \frac{\varrho_2^c(x)}{x}  + 
\frac{\varrho_4^c(x)}{x^2M^2} \right)
        \exp{ \left( - \frac{\bar x Q^2}{x M^2}  + \frac{x 
m_N^2}{M^2}\right)}\,
+
\frac{\varrho_4^c(x_0)\,e^{-(s_0-m_N^2) /M^2} }{Q^2 + x_0^2 m_N^2} \right],
\eeq
with
\bea{abbrev1}
\varrho_2^c(x) & = &
\left\{\widetilde{V}_{123}+ x \int_0^{\bar x} dx_1 V_3(x_i) \right\}
+\left\{ x  \int_0^{\bar x}dx_1 \left[2 A_1 + 
3A_3+V_3\right](x_i)
       - \widehat{A}_{123} +\widehat{V}_{123}  \right\},
\nn \\
\varrho_4^c(x) & = &
  \bigg\{ Q^2 \widetilde{V}_{123}+ x^2m_N^2 \widetilde{V}_{43}\bigg\}
+\bigg\{
     Q^2 \left( \widehat{A}_{123} +\widehat{V}_{123} \right)
    + x^2 m_N^2 \left[ \widehat{A}_{1345}- 2\widehat{A}_{34}+ 
\widehat{V}_{43}\right]
\nn \\ &&{}
    + 2x  m_N^2 \left( \mathcal{A}_1^{M(u)} - 
\widehat{\!\widehat{A}}_{123456} \right)
\bigg\}.
\eea
%
\end{widetext}
%
Note that in difference to Eq.~(\ref{LCSRIoffe}) this sum rule involves nucleon DAs on the r.h.s.,
not the pion-nucleon ones.

\section{Results } 
\setcounter{equation}{0}

The sum rules in Eqs.~(\ref{LCSRIoffe}), (\ref{GAIoffe}) involve nucleon and pion-nucleon DA that can be expanded
in contributions of conformal partial waves. To each order in the conformal expansion, one or several new 
nonperturbative parameters appear that can be related to matrix elements of local operators, as 
detailed in Ref.~\cite{BFMS00}. In the simplest approximation (asymptotic DAs) the sum rules depend on a single 
parameter which is the ratio of the twist-4 and the twist-3 matrix elements of the three-quark operators 
of the lowest dimension (e.g. without derivatives) that differ by the quark spin projection on the light-cone.
This ratio is known to ca. 20-30\% accuracy from the QCD sum rule calculations 
(see e.g.~\cite{Braun:2006hz}).
To the next-to-leading-order in the conformal spin, several more parameters enter that characterize the momentum 
fraction carried by a particular quark in the valence component of the nucleon wave function, both in the leading and
higher twists. These parameters have been estimated using QCD sum rules \cite{Chernyak:1984bm,King:1986wi,Chernyak:1987nu}  
but the accuracy of such estimates is suspected to be low. By this reason we prefer to use the set of parameters 
that are intermediate between the asymptotic and the QCD sum rule motivated DAs, and are 
chosen in such a way that the nucleon electromagnetic and axial form factors are described well within the light-cone sum rule 
approach, see Ref.~\cite{Braun:2006hz}. We will refer to this set of the DAs as the Braun-Lenz-Wittmann (BLW) model. 
For leading twist, these parameters are close to the model of Ref.~\cite{Bolz:1996sw}.    

In order to minimize further the uncertainty due to the choice of the DAs, the Borel parameter $M^2$ and the continuum 
threshold $s_0$, we present below the results for the ratios of the sum rules for the pion electroproduction amplitudes at threshold, 
Eqs.~(\ref{LCSRIoffe}), to the nucleon axial form factor, Eq. (\ref{GAIoffe}):
\beq{weplot1}
    R_1^{\pi N} = \sqrt{2}Q^2 G_1^{\pi N}/(G_A m_N^2)
\eeq
 and
\beq{weplot2}
    R_2^{\pi N} = \sqrt{2}G_2^{\pi N}/G_A\,.
\eeq
For small momentum transfers, the first ratio is predicted to be one for the $\pi^+n$ and zero for $\pi^0n$  
final state (\ref{softpi1}), while
the second ratio vanishes in both cases in the soft-pion limit  (\ref{softpi2}).

The LCSR  (\ref{GAIoffe}) for the axial form factor itself is studied in the range $Q^2 = 1-10$~GeV$^2$ 
in Ref.~\cite{Braun:2006hz}. The results are consistent (to 20\% accuracy) with the dipole parametrization
\bea{dipole3}  
G_A(Q^2)  & = & \frac{g_A}{\left(1+\frac{Q^2}{M_A^2} \right)^2}, 
\eea  
with $g_A=1.267\pm 0.004$ and the mass parameter $M_A = 1$~{GeV} that is suggested 
by both the neutrino scattering and pion electroproduction data at low $Q^2$.  

\begin{figure}
  \includegraphics[width=0.40\textwidth,angle=0]{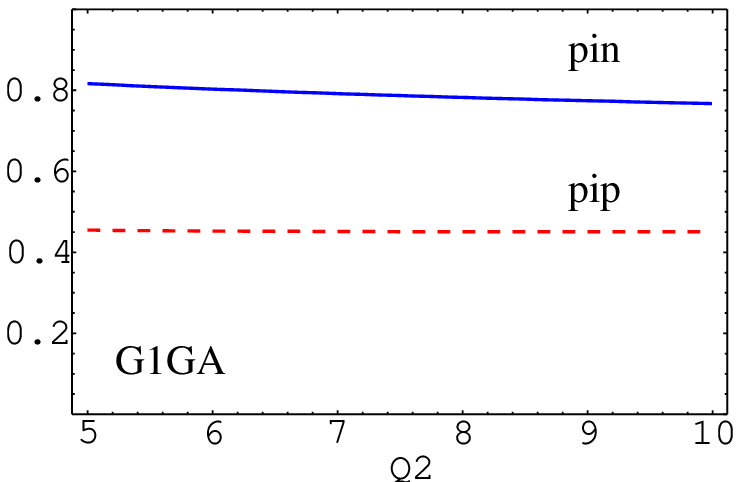}\\[2mm]
  \includegraphics[width=0.40\textwidth,angle=0]{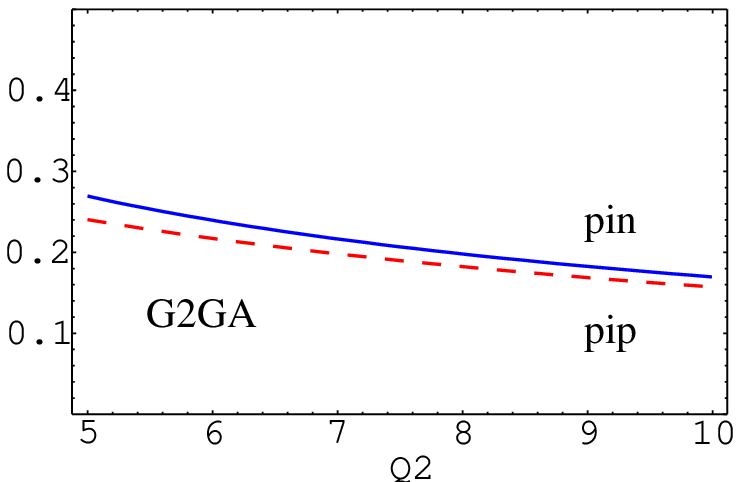}
\caption{Pion electroproduction amplitudes at threshold, (\ref{def:gff}) normalized to the 
 nucleon axial form factor: $\pi^+ n$ (solid blue curves) and $\pi^0 p$ (dashed red curves).
 The calculation is done using the BLW model \cite{Braun:2006hz} for the nucleon DAs.
The color identification refers to the on-line version.}
\label{fig:results}
\end{figure}
 
The results are shown in Fig.~\ref{fig:results} for the range of momentum transfers $Q^2 = 5-10$~GeV$^2$.
This range is chosen because for lower momenta the contribution of the pion-nucleon intermediate state 
becomes large and must be subtracted, and, on the other hand, for large momenta the (uncalculated) 
radiative corrections ${\mathcal O}(\alpha_s(M))$ are expected to play an increasing role. 

It attracts attention that the $R_1$ ratio is almost constant in the whole $Q^2$ range, i.e. the $Q^2 G_1^{\pi N}$
amplitude turns out to be proportional to the axial form factor of the nucleon. This proportionality holds 
for all models of the nucleon DAs that we are considering, whereas the value of $R_1$ does depend on the shape
of the DAs. We obtain
\bea{R1}
  R_1^{\pi^+ n} &=&  0.80\pm 0.10\,,
\nonumber\\
  R_1^{\pi^0 p} &=&  0.45\pm 0.08\,,
\eea  
where the given error estimate corresponds to the difference in the sum rule predictions using BLW and asymptotic 
DAs. 

The second amplitude, $G_2^{\pi N}$, comes out similar in magnitude but 
different sign for the $\pi^+n$ and $\pi^0 p$ final states.
The sum rule results for the $R_2^{\pi N}$ ratio, shown in Fig.~\ref{fig:results} can be 
parametrized as 
\beq{R21}
   R_2^{\pi N}(Q^2) = \frac{R_2^{\pi N}(Q_0^2)}{1+ b (Q^2-Q_0^2)}
\eeq
with the reference scale $Q_0^2=7.5$~GeV$^2$, the slope $b \simeq 0.1$~GeV$^{-2}$ and
the values 
\bea{R22}
   R_2^{\pi^+ n}(Q_0^2) &=& \phantom{-}0.21\pm 0.15 \,, 
\nonumber\\
   R_2^{\pi^0 p}(Q_0^2) &=& {-}0.19 \pm 0.07 \,, 
\eea
where the error estimates are obtained as in Eq.~(\ref{R1}).

Using our central values for the amplitudes $G_{1,2}^{\pi N}$ we obtain the 
contribution to the structure function $F_2$ (\ref{DIS}) which in the sum of all
terms is almost the same as the corresponding contribution in the soft pion limit,
cf. Eqs.~(\ref{softpi1}), (\ref{softpi2}).
This is in disagreement to Ref.~\cite{PPS01} where a much larger structure 
function was obtained. The inspection shows that our result for the $\pi^+n$ 
production coincides with that of \cite{PPS01} within the errors, while 
our prediction for the $\pi^0 p$ cross section is smaller by an order 
of magnitude. As the result, the DIS cross section obtained in \cite{PPS01} is 
completely dominated (90\%) by the contribution of the $\pi^0 p$ state
(compare Eqs.~(11) and (12) in \cite{PPS01}),
whereas in our calculation the $\pi^0$ to $\pi^+$ production ratio is only of 
order 1/3. Although the precise numbers are sensitive to the models of the nucleon DAs,
there seems to be a qualitative difference that would be interesting to check experimentally.

\section{Summary and Conclusions} 
\setcounter{equation}{0}

We have given an analysis of the pion electroproduction amplitudes at threshold
in the light-cone sum rule approach, using the methods of current algebra 
to construct the complete set of pion-nucleon distribution amplitudes in the 
soft-pion limit. The sum rule derived in this work are tree-level and may 
be affected by large (but calculable) radiative corrections. Having this in mind,
the numerical results presented in Section~5 have to be considered as semi-quantitative.
Our calculation suggests that the $\pi^0$ to $\pi^+$ production ratio 
becomes roughly of order 1/3 at large momentum transfers $Q^2 > 7$~GeV$^2$, 
and the amplitudes $G_2^{\pi^0 p}$ and $G_2^{\pi^+ n}$ are not small. They are predicted to 
be similar in magnitude and have opposite sign which can be tested in experiment
using polarization transfer techniques. Working out the concrete experimental 
predictions (e.g. corresponding asymmetries) goes beyond the tasks of this paper and 
will be considered separately.

Main assumption underlying the present calculation is that the matching of the QCD 
operator expansion with the dispersion relation in terms of hadronic states can be 
done at sufficiently large scales where pions do not present independent degrees 
of freedom. In practice, the relevant values of the Borel parameter are close to the 
cutoff scale of the chiral perturbation theory, so this assumption may need a
further check and better justification. Another direction of further work is obviously the 
inclusion of radiative corrections to the sum rules, which are enhanced at large
momentum transfers. For practical applications one may need to add contributions of pole
terms and generalize our analysis of the pion-nucleon distribution amplitudes to small 
off-threshold pion momenta by using the Omnes representation.

\section*{Acknowledgements} 
\setcounter{equation}{0}
V.B. thanks M.~V.~Polyakov for the discussion that initiated our interest to this 
problem and useful comments. The work of D.I. was partially supported by grants 
RFBR-05-02-16211, NSh-5362.2006.2 and by the Helmholtz Association, grant VH-NG-004. 
The work by A.P. was supported by the Studienstiftung des deutschen Volkes. 

%
\begin{widetext}
%

\appendix  
\renewcommand{\theequation}{\Alph{section}.\arabic{equation}}  
\section*{Appendices}

\section{Pion-Nucleon GDAs of Higher  Twists } 
\setcounter{equation}{0}

In this Appendix we give without derivation the expressions for three-quark pion-nucleon generalized distribution 
amplitudes of higher twist.

\subsection{$n\pi^+$ distribution amplitudes}
Definition of the GDAs follows Eq.~(\ref{npi+}). For the twist-4 we obtain:
\begin{eqnarray}
V_2^{n\pi^+}(1,2,3)&=&  -\frac{1}{\sqrt{2}}\Bigg\{
\frac{1}{2}\Bigl(\Big\{A_3^n(1,3,2)-P_1^n(1,3,2)+S_1^n(1,3,2)-T_3^n(1,3,2)
\nonumber\\
&&-T_7^n(1,3,2)-V_3^n(1,3,2)\Big\}+\Big\{1\leftrightarrow 2\Big\}\Bigr)
\Bigg\}\,,
\nonumber\\     
~
A_2^{n\pi^+}(1,2,3)&=&   -\frac{1}{\sqrt{2}}\Bigg\{
\frac{1}{2}\Bigl(\Big\{A_3^n(1,3,2)+P_1^n(1,3,2)-S_1^n(1,3,2)+T_3^n(1,3,2)
\nonumber\\&&
+T_7^n(1,3,2)-V_3^n(1,3,2)\Big\}-\Big\{1\leftrightarrow 2\Big\}\Big)
\Bigg\}\,, 
\nonumber\\     
~
V_3^{n\pi^+}(1,2,3)&=&   \frac{1}{\sqrt{2}}\Bigg\{
\frac{1}{2}\Bigl(\Big\{A_2^n(1,3,2)-P_1^n(1,3,2)+S_1^n(1,3,2)+T_3^n(1,3,2)
\nonumber\\&&
+T_7^n(1,3,2)+V_2^n(1,3,2)\Big\}+\Big\{1\leftrightarrow 2\Big\}\Bigr)
\Bigg\}\,, 
\nonumber\\
~
A_3^{n\pi^+}(1,2,3)&=&   \frac{1}{\sqrt{2}}\Bigg\{
\frac{1}{2}\Bigl(\Big\{-A_2^n(1,3,2)-P_1^n(1,3,2)+S_1^n(1,3,2)+T_3^n(1,3,2)
\nonumber\\&&
+T_7^n(1,3,2)-V_2^n(1,3,2)\Big\}-\Big\{1\leftrightarrow 2\Big\}\Bigr)
\Bigg\}\,,
\nonumber\\   
~
   S_1^{n\pi^+}(1,2,3)&=&   -\frac{1}{\sqrt{2}}\Bigg\{
\frac{1}{4}\Bigl(\Big\{A_2^n(1,3,2)+A_3^n(1,3,2)+P_1^n(1,3,2)+S_1^n(1,3,2)
\nonumber\\&&
+2T_2^n(1,3,2)+T_3^n(1,3,2)-T_7^n(1,3,2)-V_2^n(1,3,2)
\nonumber\\&&
+V_3^n(1,3,2)\Big\}-\Big\{1\leftrightarrow 2\Big\}\Bigr)
\Bigg\}\,,
\nonumber\\
~
P_1^{n\pi^+}(1,2,3)&=& -\frac{1}{\sqrt{2}}\Bigg\{
\frac{1}{4}\Bigl(\Big\{-A_2^n(1,3,2)-A_3^n(1,3,2)+P_1^n(1,3,2)+S_1^n(1,3,2)
\nonumber\\&&
+2T_2^n(1,3,2)+T_3^n(1,3,2)-T_7^n(1,3,2)+V_2^n(1,3,2)
\nonumber\\&&
-V_3^n(1,3,2)\Big\}-\Big\{1\leftrightarrow 2\Big\}\Bigr)
\Bigg\}\,,
\nonumber\\
~
T_2^{n\pi^+}(1,2,3)&=& -\frac{1}{\sqrt{2}}\Bigg\{
\frac{1}{2}\Bigl(\Big\{P_1^n(1,3,2)+S_1^n(1,3,2)-T_3^n(1,3,2)+T_7^n(1,3,2)\Big\}
\nonumber\\&&
+\Big\{1\leftrightarrow 2\Big\}\Bigr)
\Bigg\}\,,
\nonumber\\
~
T_3^{n\pi^+}(1,2,3)&=& \frac{1}{\sqrt{2}}\Bigg\{
\frac{1}{4}\Bigl(\Big\{A_2^n(1,3,2)-A_3^n(1,3,2)-P_1^n(1,3,2)
-S_1^n(1,3,2)\nonumber\\&&
+2T_2^n(1,3,2)-T_3^n(1,3,2)+T_7^n(1,3,2)-V_2^n(1,3,2)
\nonumber\\&&
-V_3^n(1,3,2)\Big\}+\Big\{1\leftrightarrow 2\Big\}\Bigr)
\Bigg\}\,,
\nonumber\\
~
T_7^{n\pi^+}(1,2,3)&=& \frac{1}{\sqrt{2}}\Bigg\{
\frac{1}{4}\Bigl(\Big\{A_2^n(1,3,2)-A_3^n(1,3,2)+P_1^n(1,3,2)
+S_1^n(1,3,2)\nonumber\\&&-2T_2^n(1,3,2)
+T_3^n(1,3,2)-T_7^n(1,3,2)-V_2^n(1,3,2)
\nonumber\\&&-V_3^n(1,3,2)\Big\}+\Big\{1\leftrightarrow 2\Big\}\Bigr)
\Bigg\}.
\end{eqnarray}
The expressions for twist-5 GDAs are identical up to the substitution
\begin{eqnarray}\label{Ersetzung5}
\lefteqn{\{S_1^{n\pi^+}, P_1^{n\pi^+}, V_2^{n\pi^+}, V_3^{n\pi^+}, A_2^{n\pi^+}, A_3^{n\pi^+}, T_2^{n\pi^+}, T_3^{n\pi^+}, 
    T_7^{n\pi^+}\} \rightarrow}
\nonumber\\
&&\hspace{3cm}\rightarrow \{S_2^{n\pi^+},P_2^{n\pi^+},V_5^{n\pi^+},V_4^{n\pi^+},A_5^{n\pi^+},A_4^{n\pi^+},T_5^{n\pi^+},T_4^{n\pi^+},
    T_8^{n\pi^+}\},
\end{eqnarray}
and the similar replacement of the neutron DAs on the r.h.s.

Finally, the expressions for twist-6 GDAs are identical to twist-3 with the substitution
\begin{eqnarray}\label{Ersetzung6}
\{V_1^{n\pi^+},A_1^{n\pi^+},T_1^{n\pi^+}\}&\rightarrow &\{V_6^{n\pi^+},A_6^{n\pi^+},T_6^{n\pi^+}\}
\end{eqnarray}
and the similar substitution of the neutron DAs.

\subsection{$p\pi^0$ distribution amplitudes}
Definition of the GDAs follows Eq.~(\ref{npi+}). For twist-4 we obtain
\begin{eqnarray}
  S_1^{p\pi^0}(1,2,3)&=&-\frac{1}{2}\Big\{2P_1^p(1,3,2)-S_1^p(1,2,3)\Big\},
\nonumber\\
  P_1^{p\pi^0}(1,2,3)&=& -\frac{1}{2}\Big\{2S_1^p(1,3,2)-P_1^p(1,2,3)\Big\},
\nonumber\\
 V_2^{p\pi^0}(1,2,3) &=& \phantom{-}\frac{1}{2}V_2^{p}(1,2,3)\,, \qquad V_3^{p\pi^0}(1,2,3)=\frac{1}{2}V_3^p(1,2,3)\,,
\nonumber\\
 A_2^{p\pi^0}(1,2,3) &=& \phantom{-}\frac{1}{2}A_2^{p}(1,2,3)\,, \qquad A_3^{p\pi^0}(1,2,3)=\frac{1}{2}A_3^p(1,2,3)\,,
\nonumber\\
 T_2^{p\pi^0}(1,2,3) &=& -\frac{1}{2}T_2^p(1,2,3)\,, \qquad T_3^{p\pi^0}(1,2,3)=\frac{1}{2}\Big\{T_3^p(1,2,3)+2T_7^p(1,2,3)\Big\}\,,
\nonumber\\
 T_7^{p\pi^0}(1,2,3) &=& \frac{1}{2}\Big\{T_7^p(1,2,3)+2T_3^p(1,2,3)\Big\}.
\label{ppi0}
\end{eqnarray}
The expressions for twist-5 and twist-6 GDAs are identical to those for twist-4 and twist-3, respectively, 
with obvious substitutions as described for the $n\pi^+$ case.

\subsection{$p\pi^+$ distribution amplitudes}
{}For completeness we present here the expressions for the $p\pi^+$ GDAs. 
They are defined as the corresponding matrix element of the trilocal light-ray operator 
\begin{eqnarray}
\hat{O}^{uuu}_{\al\ga\be}=\epsilon^{ijk}u_\al^i(a_1z)u_\be^j(a_2z)u_\ga^k(a_3z)
\end{eqnarray}
and can be relevant, e.g. for weak form factors corresponding to the charged $W$-boson exchange.
Working out the necessary commutators we obtain
 \begin{eqnarray}
{}  [Q^1_5,\hat{O}_{\al\be\ga}^{uuu}(z)] &=&  
   -\frac{1}{2}\Big\{(\gamma_5)_{\al \lambda}\hat{O}^{duu}_{\lambda\beta\ga}(z)
     + (\gamma_5)_{\be \lambda}\hat{O}^{udu}_{\alpha \lambda\ga}(z)
     + (\gamma_5)_{\gamma \lambda}\hat{O}^{uud}_{\alpha\beta \lambda}(z)\Big\}\,, 
\nonumber\\
{}  [Q^2_5,\hat{O}_{\al\be\ga}^{uuu}(z)] &=& 
    \phantom{-}\frac{i}{2}\Big\{(\gamma_5)_{\alpha \lambda} \hat{O}^{duu}_{\lambda\beta\ga}(z)
     + (\gamma_5)_{\beta \lambda} \hat{O}^{udu}_{\alpha \lambda\ga}(z)
     + (\gamma_5)_{\gamma \lambda} \hat{O}^{uud}_{\alpha\beta \lambda}(z)\Big\}\,,
\end{eqnarray}
 In fact, a separate calculation is not needed in this case. For all GDAs:
\begin{eqnarray}
 S_i^{p\pi^+}(1,2,3) &=& -S_i^{n\pi^+}(1,2,3)+S^p_i(1,2,3)\,,
\nonumber\\
 P_i^{p\pi^+}(1,2,3) &=& -P_i^{n\pi^+}(1,2,3)+P^p_i(1,2,3)\,,
\nonumber\\
 V_i^{p\pi^+}(1,2,3) &=& \phantom{-} V_i^{n\pi^+}(1,2,3)+V^p_i(1,2,3)\,,
\nonumber\\
 A_i^{p\pi^+}(1,2,3) &=& \phantom{-} A_i^{n\pi^+}(1,2,3)+A^p_i(1,2,3)\,,
\nonumber\\
 T_i^{p\pi^+}(1,2,3) &=& -T_i^{n\pi^+}(1,2,3)+T^p_i(1,2,3)\,.
\end{eqnarray}

\section{$x^2$-corrections}
\label{app:b}
\subsection{$\V_1^{\pi N,M}$ corrections}

The method to calculate the $\mathcal{O}(x^2)$ contributions in the OPE of 
the baryon operators
in a diquark-quark configuration (i.e. when the light-cone positions of 
two quarks coincide) is described in detail
in  Refs.~\cite{Braun:2001tj,Braun:2006hz} and need not be repeated here.
The functions $\mathcal{V}^{M}$ can be found as solutions of the following 
equations for the moments:
\begin{eqnarray}
\int \dd  x_3 \,x_3^n \,\V_1^{\pi N,M(d)}(x_3)
&=&
\frac{1}{(n+1)(n+2)} \bigg[(- 2 V_1 + V_3 + V_4 + 2 V_5)^{(d)(n+2)}
\bigg]^{\pi N}
\\  &&{} +
\frac{1}{(n+1)(n+3)} \left[ (n+3) V_1^{(d)(n+2)}
+ (V_1 -V_2)^{(d)(n+2)} - (V_1 +V_5)^{(d)(n+1)}\right]^{\pi N}\!\!,
\nn\\
\int \dd x_2  \,x_2^n \,\V_1^{\pi N,M(u)}(x_2)
                &=& \frac{1}{(n+1)(n+3)} \left[ (n+3) V_1^{(u)(n+2)}
+(V_1 -V_2)^{(u)(n+2)} \right]^{\pi N}
\nn \\ &&
  +\frac{1}{(n+1)(n+2)} \bigg[(- 2 V_1 + V_3 + V_4 + 2 V_5)^{(u)(n+2)}
\bigg]^{\pi N}\!\!.
\end{eqnarray}
We present the solutions as
\begin{eqnarray}
\V_1^{\pi N,M(u)}=\frac{1}{24}\Big\{\la_1C_{\la}^{u,\pi N}+f_N 
C_{f}^{u,\pi N }\Big\}~,
\nn \\
\V_1^{\pi N,M(d)}=\frac{1}{24}\Big\{\la_1C_{\la}^{d,\pi N}+f_N 
C_{f}^{d,\pi N }\Big\}~,
\end{eqnarray}
where
\begin{eqnarray}
C_{\la}^{u,p\pi^0}&=&-\frac{1}{2}(1-x_2)^3\Big\{6 + 18x_2 + 19x_2^2+ 
10f_1^ux_2^2 - 3(9 + 10f_1^u)x_2^3
- 4f_1^d(1 +3x_2+ 26x_2^2 - 25x_2^3)\Big\},
\nn \\
C_{f}^{u,p\pi^0 }&=&\frac{1}{2}(1-x_2)^3\Big\{6 + 18x_2+ 143x_2^2 + 
297x_2^3 - 408x_2^4 - 10A_1^u(1 - 3x_2)x_2^2
\nn \\&& -
2V_1^d(4 +12x_2  +7x_2^2 + 599x_2^3 - 612x_2^4)\Big\},
\nn \\
C_{\la}^{d,p\pi^0}&=& - \frac{1}{2} \Big\{3x_3^2(1-x_3)(87 -33x_3 -13x_3^2 
+ 11x_3^3) + 156 x_3^2 \ln[x_3]
\nn \\&&{} +
2f_1^d \Big[(1-x_3)(4 + 4 x_3 - 351 x_3^2 + 169 x_3^3 + 49 x_3^4 - 55 
x_3^5) - 180 x_3^2 \ln[x_3]\Big] \Big\},
\nn \\
C_{f}^{d,p\pi^0 }&=&-\frac{1}{2}\Big\{x_3^2(1-x_3)(1979 - 877x_3 - 
1513x_3^2 + 2232x_3^3 - 816x_3^4)
+996x_3^2 \ln[x_3]
\nn \\&&{}-
4V_1^d \Big[(1-x_3)(4 + 4 x_3 + 1527 x_3^2 - 893 x_3^3 -923 x_3^4 + 1613 
x_3^5 -612 x_3^6)+720x_3^2\ln[x_3]\Big]\Big\},
\nn \\
C_{\la}^{u,n\pi^+}&=&-\frac{1}{\sqrt{2}}(1-x_2)^3\Big\{2  +6x_2 - 75x_2^2 
+ 43x_2^3
+ 4f_1^u(1 +3x_2 +26x_2^2 - 25x_2^3)
\nn \\&&
{}+ 2f_1^d(4 + 12x_2 + 89x_2^2 - 55x_2^3)\Big\},
\nn \\
C_{f}^{u,n\pi^+}&=&-\frac{1}{\sqrt{2}}(1-x_2)^3 \Big\{10 +30x_2 + 425x_2^2 
- 337x_2^3 - 30V_1^d(1 - 3x_2)x_2^2
\nn \\&&{}-
2A_1^u(4 +12x_2 +17x_2^2 +569x_2^3 - 612x_2^4)\Big\},
\nn \\
C_{\la}^{d,n\pi^+}&=& 
\frac{1}{\sqrt{2}}\Big\{(1-x_3)(-8-8x_3+441x_3^2-239x_3^3-59x_3^4+77x_3^5)+204 
x_3^2\ln[x_3]
\nn\\&&{}
+2(f_1^u+2f_1^d)\Big[
(1-x_3)(4+4x_3-351x_3^2+169x_3^3+49x_3^4-55x_3^5)-180x_3^2\ln[x_3]\Big] 
\Big\},
\nn \\
C_{f}^{d,n\pi^+}&=&-\frac{1}{\sqrt{2}}\Big\{
(1-x_3)(16+16x_3+171x_3^2-941x_3^3+847x_3^4-217x_3^5) - 108 x_3^2 \ln[x_3]
   \nn \\&&{}+
4A_1^u\Big[(1-x_3)(4+4x_3+1527x_3^2-893x_3^3-923x_3^4+1613x_3^5-612x_3^6)
  + 720 x_3^2 \ln[x_3] \Big] \Big\}.
\end{eqnarray}

\subsection{$\A_1^{\pi N,M}$ corrections}

Similarly we get the equations for the $\mathcal{A}^{M}$ functions:
\begin{eqnarray}
\int \dd  x_3 \,x_3^n \,\A_1^{\pi N,M(d)}(x_3)
&=&
\frac{1}{(n+1)(n+2)} \bigg[(- 2 A_1 + A_3 + A_4 + 2 A_5)^{(d)(n+2)}
\bigg]^{\pi N}
\\  &&{} +
\frac{1}{(n+1)(n+3)} \left[ (n+3) A_1^{(d)(n+2)}
+ (A_1 -A_2)^{(d)(n+2)} - (A_1 +A_5)^{(d)(n+1)}\right]^{\pi N}\!\!,
\nn\\
\int \dd x_2  \,x_2^n \,\A_1^{\pi N,M(u)}(x_2)
                &=& \frac{1}{(n+1)(n+3)} \left[ (n+3) A_1^{(u)(n+2)}
+(A_1 -A_2)^{(u)(n+2)} \right]^{\pi N}
\nn \\ &&
  +\frac{1}{(n+1)(n+2)} \bigg[(- 2 A_1 + A_3 + A_4 + 2 A_5)^{(u)(n+2)}
\bigg]^{\pi N}
\!\!,
\end{eqnarray}
We obtain
\begin{eqnarray}
{\cal A}_1^{\pi N,M(d)}&=&0\,,
\nonumber\\
\A_1^{\pi N,M(u)}&=&\frac{1}{24}(1-x_2)^3 \Big\{\la_1D_{\la}^{u,\pi N}+f_N 
D_{f}^{u,\pi N }\Big\}.
\end{eqnarray}
with the coefficients
\begin{eqnarray}
D_{\la}^{u,p\pi^0}&=&-\frac{1}{2}\Big\{
x_2^2(29-45 x_2) +2 f_1^u(4+12 x_2-51 
x_2^2+45x_2^3)-4f_1^d(2+6x_2+17x_2^2-30x_2^3)\Big\},
\nn \\
D_{f}^{u,p\pi^0 }&=& -\frac{1}{2}\Big\{
8+24 x_2 -21 x_2^2 +2 A_1^u(4 +12x_2- 9x_2^2 -609x_2^3 + 612x_2^4)+45 
x_2^3 +10 V_1^d x_2^2(1-3x_2) \Big\},
\nn \\
D_{\la}^{u,n\pi^+}&=&\frac{1}{\sqrt{2}}\Big\{
3 x_2^2(47-55x_2)-4f_1^u(2+6x_2+17x_2^2-30x_2^3)+2f_1^d(4+12 x_2-221 
x_2^2+255x_2^3)\Big\},
\nn \\
D_{f}^{u,n\pi^+}&=&\frac{1}{\sqrt{2}}\Big\{
16+48x_2-35x_2^2+1323x_2^3-1224x_2^4
+2A_1^u(8+24x_2-13x_2^2-1233x_2^3+1224x_2^4)
\nn\\ &&{}
+6V_1^d(4+12x_2-9x_2^2-609x_2^3+612x_2^4)
\Big\}.
\end{eqnarray}

\end{widetext}

\end{document}